\documentstyle[12pt,fleqn,times]{article}
\renewcommand{\appendix}{
  \setcounter{section}{0}
  \setcounter{subsection}{0}
  \setcounter{equation}{0}
    \def\thesection{\Alph{section}}
    \def\theequation{\thesection.\arabic{equation}}}
\begin{document}

\begin{center}

\LARGE{\bf DOUBLE PION PHOTOPRODUCTION ON THE NUCLEON: STUDY OF THE
       ISOSPIN CHANNELS}

\vspace{0,2cm}
\end{center}
\vspace{2cm}
\begin{center}
{\large{J. A. G\'{o}mez Tejedor and E. Oset}}
\end{center}
\begin{center}
{\small{Departamento de F\'{\i}sica Te\'{o}rica and IFIC, Centro Mixto
Universidad de Valencia - CSIC|| 46100 Burjassot (Valencia), Spain}}
\end{center}
\vspace{2cm}

\begin{abstract}
{\small{
A model for the $\gamma p \rightarrow \pi^+ \pi^- p$
reaction developed earlier is extended to account for all isospin channels.
The model includes $N$, $\Delta(1232)$, $N^*(1440)$ and $N^*(1520)$
as intermediate baryonic states and the $\rho$-meson as an intermediate
$2 \pi$ resonance.
Although many terms contribute to the cross section, some channels exhibit
particular sensitivity to certain mechanisms of resonance excitation or
decay and the reactions provide novel information on such mechanisms.
In particular the $\gamma N \rightarrow N^*(1520) \rightarrow \Delta \pi$
 process affects all the channels and is a key ingredient in the
interpretation of the data.
Comparison is made with all available data and the agreement is good in some
channels.
The remaining discrepancies in some other channels are discussed.
}}
\end{abstract}
\newpage
\section{Introduction.}

There are three possible double pion photoproduction reactions on
the proton, and three on the neutron:

\begin{equation}                            \label{reaction}
\begin{array}{rlcrl}
 \hspace{1cm} {\it (a)} & \gamma p \rightarrow \pi^+ \pi^- p & \hspace{1cm} &
 \hspace{1cm} {\it (d)} & \gamma n \rightarrow \pi^+ \pi^- n \\
 \hspace{1cm} {\it (b)} & \gamma p \rightarrow \pi^+ \pi^0 n & \hspace{1cm} &
 \hspace{1cm} {\it (e)} & \gamma n \rightarrow \pi^- \pi^0 p \\
 \hspace{1cm} {\it (c)} & \gamma p \rightarrow \pi^0 \pi^0 p & \hspace{1cm} &
 \hspace{1cm} {\it (f)} & \gamma n \rightarrow \pi^0 \pi^0 n \\
\end{array}
\end{equation}

These reactions have been extensively studied experimentally in the past
(\cite{abbhhm,gialanella,carbonara,piazza,cambridge,crouch,hilpert}).
New improvements in
experimental techniques and facilities have reopened the study of these
reactions at Mainz, with two experiments on the proton
\cite{daphne,taps}.

{}From the theoretical point of view, only the reaction ${\it (a)}$ has been
studied with one early model \cite{luke} which considers only
5 Feynman diagrams, and a more complete one \cite{yo,tesina} which includes
$N$, $\Delta (1232)$, $N^* (1440)$ and $N^* (1520)$ as intermediate
baryonic states and the $\rho$-meson as intermediate $2 \pi$-resonance.
This model reproduces fairly well the experimental cross sections below
$E_{\gamma} = 800 \, MeV$, and the invariant mass distributions
even at higher energies \cite{yo}.

On the other hand, the double pion photoproduction reactions
on nucleons has been recently
 studied at threshold from the point of view of chiral perturbation
theory \cite{benmerrouche,meisner,meis,ben}.
In ref. \cite{meisner} the authors show that due to finite chiral loops the
cross
section at threshold for final states with two neutral pions is considerably
enhanced.

Thus, with these reactions becoming a target of new experimental and
theoretical studies, plus
interesting medium effects predicted for the $(\gamma , \pi^+  \pi^- )$
reaction in nuclei \cite{manolo}  (in analogy to those already found for
the $(\pi , 2 \pi)$ reaction \cite{pidospi,grion}), a thorough
theoretical study of the $\gamma N \rightarrow \pi \pi N$ reaction
is timely and opportune.

In this work, we have improved the model of reference \cite{yo},
and have extended it to the other isospin channels of Eq. (\ref{reaction}).
Moreover, although is not the purpose of this paper to perform a thorough
analysis
of the $\gamma N \rightarrow \pi \pi N$ reaction at threshold, we have
calculated the total cross section for the $\gamma p \rightarrow
\pi^+ \pi^- p$ and  $\gamma p \rightarrow
\pi^0 \pi^0 p$ reactions at threshold energies, and compared our
result with the result given in ref. \cite{benmerrouche,meisner}
based on Chiral Perturbation Theory.

The model is based on the coupling of photons and pions to nucleons
and resonances using effective Lagrangians and thus leading to a set of
Feynman diagrams at the tree level.
As well as in ref. \cite{yo}, we do not implement unitarity in the final
states, but we made an estimate of possible uncertainties owing
to unitary corrections for the $\gamma p \rightarrow \pi^+ \pi^- p$
channel.

\section{The model for the $\gamma N \rightarrow \pi \pi N$ reaction.}

\subsection{the  $\gamma p \rightarrow \pi^+ \pi^- p$ channel.}

For the  $\gamma p \rightarrow \pi^+ \pi^- p$ isospin channel
we follow strictly the model of ref. \cite{yo}, with some slight
modifications in the $N^* (1440)$ and $N^* (1520)$ couplings in order
to reproduce the new branching ratios of the last Review of Particle
Properties \cite{pdg94} (see Appendix A of the present paper
for Lagrangians and coupling
constants), and we have also improved the $N^*(1440)$ propagator,
modifying the energy dependence of the width to account explicitly
for the $N^*(1440)$ decay into one and two pions.

We classify our diagrams in one point,
two point and three point diagrams, according to the number of vertices in
the baryonic lines (see Fig. 1).
Our basic components are pions, nucleons and nucleonic
resonances: For the baryonic components
we consider  $N$,
$\Delta (1232,J^{\pi}=3/2^{+}, I=3/2)$, $N^{*}(1440,J^{\pi}=1/2^{+},I=1/2)$
and $N^{*}(1520, J^{\pi}=3/2^{-},I=1/2)$.
The $N^*(1440)$ played an important role in the $\pi N \rightarrow \pi \pi
N$ reaction \cite{pidospi,grion} in some isospin channels and for this
reason, together with the fact that the $N^*(1440)$ excited in the $\gamma
N$ collision is on-shell for $E_{\gamma} \simeq 600$ $MeV$, we pay a special
attention to this resonance.
The $N^{*} (1520)$ has a particularly large
coupling to the photons and proves to be an important ingredient, mostly
due to its interference with the dominant term of the
process, the $\gamma N \rightarrow \Delta \pi$ transition through the
gauge Kroll-Ruderman term. Higher resonances have a weaker coupling to
photons
and do not interfere with the dominant term, hence their contribution is
small,
at least for photon energies below $800$ $MeV$, Mainz energies,
where our model is supposed to work. Because of the important coupling of the
$\rho$-meson to the
two pion system and the $\gamma \pi$ system we have also considered terms
involving the $\rho$-meson.
These terms are
only relevant at high energies but show up clearly in the two pion invariant
mass distributions at these energies \cite{yo}.

With these considerations, the basic
diagrams which we consider have the structure shown in Fig.$\:$1. In
diagram
(a) and (b) the one point $NN \pi \pi$ coupling stands for the
$s$-wave $\pi N$
interaction. We consider there only the isoscalar part of the amplitude. The
isovector part is mediated by $\rho$ exchange \cite{ericson} and hence it is
explicitly taken into account in diagrams (f) and (h).
Diagram (c) contains the gauge
term $N N \pi \gamma $ or Kroll-Ruderman term. We use a pseudovector coupling
for the $N N \pi $ vertex and this allows us to consider exclusively positive
energy intermediate states in the hadronic propagators \cite{rafa}. In the two
point and three point diagrams we include nucleon and the resonances as
intermediate states. However, while all possible diagrams with N and
$\Delta (1232)$
intermediate states are considered, we omit some with $N^{*}(1440)$
intermediate states which are very small.
For the $N^*(1520)$ intermediate states we keep only the
term which interferes
with the dominant term of the amplitude
($\Delta (1232)$ Kroll-Ruderman term). In addition, all different time
orderings of the diagrams are considered.

The diagram (g) involves a gauge term $\rho \pi \pi \gamma $ coming from
minimal
coupling in the $\rho \pi \pi$ vertex which contains a derivative coupling.
The diagram (i) involves the $\gamma \rho \pi$ vertex which appears in the
$\rho \rightarrow \gamma \pi$ decay.
Finally diagram (j) contains the anomalous coupling
$\gamma 3 \pi$ \cite{cohen}.

The Feynman diagrams considered and detailed calculations can be found
in refs. \cite{yo,tesina}.

In spite of the small contribution of the $N^*(1440)$ terms in the reaction
$\gamma p \rightarrow \pi^+ \pi^- p$ (less than 2\% \cite{yo}){\footnote
{However, for the case of photo-production of two neutral pions,
the contribution
of these terms is more relevant (between 10\% and 60\%, depending on the
energy), thus it is important to treat them carefully.}},
in order to be consistent we have revised the value of the coupling constants
involving the $N^*(1440)$ so that one may fit the new branching ratios of
ref. \cite{pdg94} (see Appendix A of the present paper
for coupling constants).
This updating gives us very small differences
compared to the values given in ref. \cite{yo}.

The main change appears in the $N^* \Delta \pi$ coupling constant due to
the fact that in the study of the decay
of $N^*(1440)$ into $\Delta \pi$
of ref. \cite{yo} we considered the $\Delta (1232)$ as a stable particle.
Now we modify this in order to take into
account the finite width of the $\Delta (1232)$.
The fact that the $\Delta (1232)$ width is not small compared
to the mass difference between the $N^*(1440)$ and the $\Delta (1232)$
makes this correction advisable.

Making use of the Lagrangian (A.7), the decay width in the c.m. system
of the $N^*(1440)$ into $\Delta \pi$, considering the $\Delta (1232)$
as a stable particle (zero width),
is given by:
$$
\Gamma_{N^* \rightarrow \Delta \pi} =
 \frac{1}{ ( 2 \pi)^2} \frac{g^2_{\Delta N^* \pi}}{\mu^2}
 \int d^3 p \frac{m_{\Delta}}{E_{\Delta}(\vec{p}\, )}
 \frac{1}{2 \omega(\vec{p}\, )} \frac{4}{3} \vec{p}\,^2
 \delta(m_{N^*} - E_{\Delta}(\vec{p}\, ) -  \omega(\vec{p}\, ) )
$$
\begin{equation}                            \label{gamma}
\end{equation}
where $\mu$, $m_{\Delta}$ and $m_{N^*}$ are the pion, $\Delta (1232)$ and
$N^*(1440)$ masses respectively;
$\vec{p}$ is the momentum of the outgoing pion;
$E_{\Delta}({\vec{p}}\, ) = \sqrt{{m^2_{\Delta}}+\vec{p} \, ^2}$ and
$\omega({\vec{p}}\, ) = \sqrt{\mu^2+\vec{p} \, ^2}$
are the $\Delta (1232)$ and pion energies.

In order to account for the finite $\Delta (1232)$ width,
we have  to replace the delta function of energy conservation in
Eq. (\ref{gamma})
by
\begin{equation}                        \label{im}
 \frac{-1}{\pi}
\, Im
 \frac{1}{m_{N^*} - E_{\Delta}(\vec{p}\, ) -  \omega(\vec{p}\, ) +
    \frac{i}{2}\Gamma_{\Delta}(\sqrt{s_{\Delta}})}
\end{equation}
where $\Gamma_{\Delta}(\sqrt{s_{\Delta}})$ is the delta decay width,
and $\sqrt{s_{\Delta}}$ is the $\Delta$ invariant mass, $\sqrt{s_{\Delta}}
=  \sqrt{{p^0_{\Delta}}^2 - {\vec{p}_{\Delta}}\, ^2}$
($p_{\Delta} = (p^0_{\Delta},\vec{p}_{\Delta})$ is the $\Delta$
four momentum).
With this modification,
and using for $\Gamma_{\Delta}$ the expression (B.32) of ref. \cite{yo},
we get $ g_{\Delta N^* \pi} = 2.07 $
after fitting the $N^*(1440)$ decay width into $\Delta \pi$ to the
average experimental value ($87.5$ $MeV$)\cite{pdg94}.

On the other hand, in the $N^*(1440)$ propagator
\begin{equation}                     \label{prop}
  \frac{1}{\sqrt{s_{N^*}} -m_{N^*} +
              \frac{i}{2}\Gamma_{N^*}(\sqrt{s_{N^*}})}
  \frac{m_{N^*}}{E_{N^*}(\vec{p}_{N^*})}
\end{equation}

\noindent
where
$
\sqrt{s_{N^*}} = \sqrt{{p^0_{N^*}}^2 - {\vec{p}_{N^*}}\, ^2}
$
is the $N^*(1440)$ invariant mass ($p_{N^*} =
(p^0_{N^*},\vec{p}_{N^*})$ is the $N^*(1440)$
four momentum), and $\Gamma_{N^*}(\sqrt{s_{N^*}})$
is the $N^*(1440)$ decay width.

We have included in the $N^*(1440)$ decay width, both the decay into $N \pi$
and $N \pi \pi$, thus
\begin{equation}
\Gamma_{N^*} \left( \sqrt{s_{N^*}} \right) = \Gamma_{N^* \rightarrow N \pi}
                                             \left( \sqrt{s_{N^*}} \right)
       \: + \: \Gamma_{N^* \rightarrow N \pi \pi}
                 \left( \sqrt{s_{N^*}} \right)
\end{equation}
\noindent
where $\Gamma_{N^* \rightarrow N \pi}(\sqrt{s_{N^*}})$
is the $N^*(1440)$ decay width into $N \pi$ which is given by
\begin{equation}
\Gamma_{N^* \rightarrow N \pi} \left( \sqrt{s_{N^*}} \right) =
\frac{3}{2 \pi} \left( \frac{\tilde{f}}{\mu} \right)^2
\frac{m}{m_{N^*}} | \vec{q}\,|^3 \theta (\sqrt{s_{N^*}} -m -\mu)
\end{equation}
where $\tilde{f}$ is the $N^*(1440) N \pi$ coupling constant (see Appendix
A), and $|\vec{q}\,|$ is the pion momentum in the $N^*(1440)$ centre of mass
system. We take a fraction of 65\% for the decay into $N \pi$ \cite{pdg94}.
$ \Gamma_{N^* \rightarrow N \pi \pi}(\sqrt{s_{N^*}})$
is the $N^*(1440)$ decay width into $N \pi \pi$, which we approximate
by a constant, multiplied by three body phase space.
This constant is fitted in order to reproduce the experimental $N^*(1440)$
decay width into $N \pi \pi$.
Although part of the $N^* \rightarrow N \pi \pi$ decay goes through the
$\Delta
\pi$ channel (and this is explicitly considered in the Feynman diagrams) we
follow the above prescription solely for the purpose of providing the energy
dependence of the $N^*(1440)$ width.

Apart from these small changes in the  $N^* (1440)$ terms,
the main improvement of the model of ref. \cite{yo}
is coming from the modification of the $N^* (1520) \Delta \pi$ coupling.
In ref. \cite{yo} we took the simplest Lagrangian compatible with the
conservation
laws:

\begin{equation}                \label{f}
{\cal L}_{N'^* \Delta \pi} =
i \tilde{f}_{N'^* \Delta \pi} \overline{\Psi}_{N'^*} \phi^{\lambda}
T^{\lambda}
{\Psi}_{\Delta} \: + \: h.c.
\end{equation}
where ${\Psi}_{N'^*}$, $\phi^{\lambda}$ and ${\Psi}_{\Delta}$
stand for the $N^*(1520)$, pion and $\Delta (1232)$ fields
($N'^*$ stands for the $N^*(1520)$ in the formulae from now on);
$T^{\lambda}$ is the transition isospin operator from $1/2$ to $3/2$ with
the normalization in terms of a Clebsch-Gordan coefficient
\begin{equation}                          \label{norma}
\langle \frac{3}{2}, M  \left| T^{\dagger}_{\nu} \right|
\frac{1}{2}, m \rangle
= C \left( \frac{1}{2}, 1 ,\frac{3}{2} ; m, \nu, M \right)
\end{equation}
and $\tilde{f}_{N'^* \Delta \pi}$  is the coupling constant, that we fixed
from the data of $N^* (1520)$ decay into $\Delta \pi$ of ref. \cite{pdg92}.
The sign however was chosen such as to
have constructive interference below the $N^*(1520)$ pole.
With the chosen sign the agreement with the
data was relatively good, while with the opposite sign the discrepancies were
about a factor of two and the qualitative features of the experiment were not
reproduced \cite{yo}.

Here, it is important to note that the $N^* (1520)$ can decay into $\Delta \pi$
both in $s$-wave and $d$-wave \cite{pdg94}. However, the Lagrangian of Eq.
(\ref{f})
gives only the decay into $s$-wave, not $d$-wave{\footnote
{In the previous version of the Review of Particle Properties \cite{pdg92},
there were not branching ratios to $s$-wave and $d$-wave, and  only
the total decay into $\Delta \pi$ was given. Thus, in ref. \cite{yo}
we assumed that all the decay was through the $s$-wave, and we neglected the
$d$-wave. As the fraction of decay into the
$d$-wave is even bigger than into the $s$-wave \cite{pdg94},
corrections to the coupling of Eq. (\ref{f}) must be now implemented. The
influence of these changes in the results for the $\gamma N \rightarrow
\pi \pi N$ reaction, although not large, certainly help to improve agreement
with the data, as we shall see.}}.
Then, we have modified the Lagrangian of Eq. (\ref{f})
in order to get a term which gives
a contribution to the $d$-wave decay:

\begin{equation}                \label{fg}
{\cal L}_{N'^* \Delta \pi} =
i  \overline{\Psi}_{N'^*}
\left(
   \tilde{f}_{N'^* \Delta \pi} \: - \:
   \frac{\tilde{g}_{N'^* \Delta \pi}}{{\mu}^2}
   S_{i}^{\dagger} \partial_i \, S_{j} \partial_j
\right)
\phi^{\lambda}
T^{\lambda} {\Psi}_{\Delta}
 \: + \: h.c.
\end{equation}
which gives us the vertex contribution:
\begin{equation}                \label{hfg}
-i \delta H_{N'^* \Delta \pi} = -
\left(
   \tilde{f}_{N'^* \Delta \pi} \: + \:
   \frac{\tilde{g}_{N'^* \Delta \pi}}{{\mu}^2}
   \vec{S}^{\dagger} \cdot \vec{q} \: \vec{S} \cdot \vec{q}
\right)
T^{\lambda}
\end{equation}

\noindent
where $\vec{S}$ is the $1/2$ to $3/2$ spin transition operators with the
same normalization as Eq. (\ref{norma}) , $\mu$ is
the pion mass and $\vec{q}$ is the pion momentum in the $N^*(1520)$ rest frame.

In order to fit
the coupling constants to the experimental $N^*(1520)$
decay amplitudes to $\Delta \pi$ in
$s$- and $d$-wave \cite{pdg94,manley}, we make a partial
wave expansion \cite{pin} of the transition amplitude from a state of spin
$3/2$ and third component M, to a state of spin $3/2$ and third component M':
$$
- i \, \langle \frac{3}{2} M' | \delta H_{N'^* \Delta \pi}
    | \frac{3}{2}, M \rangle =
    A_s \, Y_0^{M-M'} (\theta, \phi) \: + \:
$$
\begin{equation}                       \label{pwa}
  A_d \, C \left(2,\frac{3}{2}, \frac{3}{2}; M-M' , M',M \right) \,
Y_2^{M-M'}(\theta , \phi)
\end{equation}
\noindent
where $Y_l^m(\theta, \phi)$ are the spherical
harmonics, and $A_s$ and $A_d$ are the $s$- and $d$-wave partial amplitudes
for the $N^*(1520)$ decay into $\Delta \pi$.
In order to compare with the experimental signs of the  amplitudes
\cite{pdg94,manley} in Eq. (\ref{pwa}) we have followed the standard
``baryon-first'' phase convention \cite{phase}. In this convention, spin
couplings in all angular momentum Clebsch-Gordan coefficients are ordered such
that the orbital angular momentum $L$ comes before the intrinsic spin $S$.
The angles in the spherical harmonics
are those of the pion referred to the baryon.
Then, making use of the coupling of Eq. (\ref{hfg}), the $A_s$ and
$A_d$ amplitudes are given by:

\begin{equation}            \label{AsAd}
\begin{array}{l}
A_s = - \sqrt{4 \pi}
     \left( \tilde{f}_{N'^* \Delta \pi} \: + \:
                 \frac{1}{3} \tilde{g}_{N'^* \Delta \pi}
\frac{\vec{q}\,^2}{\mu^2}
     \right) \\ \\
     A_d = \frac{\sqrt{4 \pi}}{3} \tilde{g}_{N'^* \Delta \pi}
\frac{\vec{q}\,^2}{\mu}
\end{array}
\end{equation}

By using Eq. (\ref{pwa}) we obtain the $N^*(1520)$ width into $\Delta \pi$:
\begin{equation}                         \label{gamma2}
\Gamma_{N'^* \rightarrow \Delta \pi} =
 \frac{1}{4 \pi} \, \frac{m_{\Delta}}{m_{N'^*}} \, q
 \left( |A_s|^2 + |A_d|^2 \right)
\end{equation}
\noindent
where $m_{\Delta}$ and $m_{N'^*}$ are the $\Delta$ and $N^*(1520)$ masses
respectively, and $q$ is the momentum of the outgoing pion
in the $N^*(1520)$ rest frame.
Now, we fit the $N^*(1520)$ partial decay width in $s$- and $d$-wave
to the experiment \cite{pdg94} (we take the average values for these partial
decays, 8.5\% for the $s$-wave, and 12\% for the $d$-wave). In addition,
we impose the ratio $A_s/A_d$ to be positive, as deduced from the
experimental analysis of the $\pi N \rightarrow \pi \pi N$ reaction
\cite{manley}.
Thus, with this last restriction, we get two different solutions for
the coupling constants which differ only in a global sign:

\begin{equation}                   \label{2sol}
\begin{array}{rll}
{\it (a)} & \hspace{1cm} \tilde{f}_{N'^* \Delta \pi} = 0.911  &
            \hspace{1cm} \tilde{g}_{N'^* \Delta \pi} =-0.552 \\
{\it (b)} & \hspace{1cm} \tilde{f}_{N'^* \Delta \pi} =-0.911 &
            \hspace{1cm} \tilde{g}_{N'^* \Delta \pi} = 0.552 \\
\end{array}
\end{equation}

Now, the $\gamma p \rightarrow \pi^+ \pi^- p $ reaction allows us to
distinguish between both solutions.
In Fig. 2 we have plotted the total cross section
for both solutions.
As we can see, only solution ${\it (a)}$ fits the experiment, while the other
one under-estimates the experimental cross section by a large amount.

This strong dependence of the total cross section on the $N^*(1520)$ coupling
was
already remarked in ref. \cite{yo}, and it is due to the important
interference between the $\Delta$ Kroll-Ruderman term (diagram $(i)$
of Fig. 3) and the $\gamma p \rightarrow N^*(1520) \rightarrow \Delta \pi$
term (diagram (p) of Fig. 3) in $s$-wave \cite{yo}.
The amplitudes for these two diagrams have exactly the same structure,
except by
the $N^*(1520)$ propagator and coupling constants (see Eq. (6) of ref.
\cite{yo}, where $\tilde{f}_{N'^* \Delta \pi}$ would be substituted by
$ \tilde{f}_{N'^* \Delta \pi} + \frac{1}{3} \tilde{g}_{N'^* \Delta \pi}
\vec{q}\,^2/\mu^2$).
Then, the $N^*(1520)$ interference term is proportional to the real part of
the $N^*(1520)$ propagator \cite{yo}. There is also a contribution to the
total cross section coming from the imaginary part of the $N^*(1520)$
propagator, and also from the $A_d$ amplitude, which do not interfere.
Thus, this interference is zero at $\sqrt{s}=m_{N'^*}$,
where the real part of the $N^*(1520)$ propagator vanishes.
Then, what matters in the interference is the value of
$A_s$ for different values of $q$ than the one from the $N^*(1520)$ on-shell.
Then, the $\gamma p \rightarrow \pi^+ \pi^- p$ reaction provides us with
novel information about the $q$ dependence of the $A_s$ amplitude, respect
to the one obtained from the analysis of the $\pi N \rightarrow \pi \pi N$
reaction \cite{manley}.

This $q$ dependence in the $s$-wave amplitude
is given by the chosen Lagrangian (on the other hand
for the $A_d$ amplitude there is no choice in the momentum
dependence exhibited
in Eq. (\ref{AsAd})).
It is, thus, possible to postulate other Lagrangians which would produce a
different $q$ dependence of the $A_s$ amplitude.
For instance, in a preliminary work reported in \cite{fb95,praga} a different
Lagrangian was investigated. There we proposed a similar Lagrangian to Eq.
(\ref{fg}), but with the $3/2$ spin operators instead of the $S^i$
transition operators. With such a Lagrangian and an appropriate choice of
parameters we were able to reproduce the experimental data, though the
ratio $A_s/A_d$ was negative.

In order to investigate the most general $q$
dependence of the amplitude, we replace $\tilde{f}_{N'^* \Delta \pi}$  by
\begin{equation}                   \label{epsilon}
\tilde{f}_{N'^* \Delta \pi}
\left( 1 \: + \: \epsilon \, \, \frac{\vec{q}\,^2 -
\vec{q}\,^2_{on-shell}}{\mu^2}
\right)
\end{equation}
where $\vec{q}$ is the momentum of the decay pion, and
$\vec{q}\,^2_{on-shell}$ is de momentum of the pion for an on-shell $N^*(1520)$
decaying into $\Delta \pi$ ($|\vec{q}_{on-shell}|$ = $228 \: MeV$).
Then, we change $\epsilon$ and compare the results to the data.
We find that, to a good
approximation, $\epsilon=0$ gives the best agreement with the data, hence
supporting the Lagrangian of Eq. (\ref{fg}).

It is interesting to remark here that this $q$ dependence of the amplitudes
coincides exactly with
the predictions of the non-relativistic constituent quark models,
although the absolute values of the amplitudes are only qualitatively given
by these models \cite{quark}.

\subsection{The others isospin channels.}

For the others isospin channels we use the same model
 as for the $\gamma p \rightarrow \pi^+ \pi^- p$ case, changing
the isospin factors and introducing some terms which are only relevant in
the case of neutral pions (although their influence in the cross sections is
small).
However, we have neglected some
Feynman diagrams which were found to be very small in \cite{yo}.
The Feynman diagrams that we
take into account are shown in Fig. 3,
where all possible charge combinations allowed by charge conservation
have to be implemented for each isospin channel.
In the case of neutral pions and neutral deltas
some of the diagrams in Fig. 3 are automatically zero.
For instance, diagrams (c) and (d) of Fig. 3 with the photon
attached to a neutral pion are zero.

Nevertheless, some remarks are necessary for the isospin channels
with neutral pions:
When in the diagrams of Fig. 4 (a) and (b) we evaluate
the contribution
from the intermediate states of negative energy we get an extra contribution
with the same structure as the Kroll-Ruderman term \cite{rafa}.
In the case of charged pions, we have neglected this contribution due to
the fact that it is very small compared with the Kroll-Ruderman
terms. However, for the neutral pions, as there is no Kroll-Ruderman
term, this contribution has to be considered and,
for a nucleon in the final state is given by \cite{rafa}:
\begin{equation}                   \label{tkr}
-i T_{\alpha} =
  e \, \frac{f}{\mu} \, C(\alpha) \, \vec{\sigma} \cdot \vec{\varepsilon}
\end{equation}
where
\begin{equation}                    \label{calfa}
 C(\alpha)= \left\{ \begin{array}{ll}
                -q^0 \left( \frac{1}{p^0 + k^0 + E(\vec{p} + \vec{k} \, )}
                \: + \:
                \frac{1}{{p'}^0 - k^0 + E(\vec{p}\, '- \vec{k} \, ) } \right)
                & for \ proton \\
		& \\
                0 & for \ neutron \\
                   \end{array}
            \right.
\end{equation}
where $(p^0,\vec{p}\, )$,  $({p'}^0,\vec{p}\, ')$,  $(q^0,\vec{q}\, )$
and  $(k^0,\vec{k}\, )$
are the four-momentum of the incoming nucleon, the outgoing nucleon,
the pion and the photon respectively.

We can write Eq. (\ref{tkr}) as:
\begin{equation}                      \label{tkrg}
-i T =
  \frac{1}{2}(1+ \langle N | \tau_3 | N \rangle)
  e \, \frac{f}{\mu} \, C \, \vec{\sigma} \cdot \vec{\varepsilon}
\end{equation}
where
\begin{equation}                    \label{calfag}
 C=
          -q^0 \left( \frac{1}{p^0 + k^0 + E(\vec{p}+\vec{k} \,)} \: + \:
               \frac{1}{{p'}^0 - k^0 + E(\vec{p}\, '-\vec{k} \, ) }
               \right)
\end{equation}
and $\tau_3$ is the third component of the isospin Pauli matrices:
\begin{equation}                     \label{tau3}
 \langle p | \tau_3 | p \rangle = 1 \\
 ; \\
 \langle n | \tau_3 | n \rangle = -1
\end{equation}
The advantage of expression (\ref{tkrg}) is that it allows us to generalize it
to the case where we have a $N^*(1440)$ or a $\Delta (1232)$ in the
 final state
by the following method:
for the $N^*(1440)$ we replace the coupling constant  $N N \pi$ ($f$) by
the coupling constant $N^* N \pi$ ($\tilde{f}$):
\begin{equation}                  \label{n1440}
-i T =
  \frac{1}{2}(1+ \langle N | \tau_3 | N \rangle)
  \, e \, \frac{\tilde{f}}{\mu} \, C \, \vec{\sigma} \cdot \vec{\varepsilon}
\end{equation}

For the $\Delta (1232)$ case we replace the coupling constant $NN\pi$ ($f$) by
the $ \Delta N \pi$ coupling constant ($f^*$)
and the $\vec{\sigma}$ and $\tau_3$ operators by the
transition spin an isospin operators from $1/2$ to $3/2$ objects (with the
normalization given by Eq. (\ref{norma})).
However, the isoscalar part does not contribute in this case. Then,
for the $\Delta N \pi^0$ coupling we have:
\begin{equation}                  \label{delta}
-i T =
  \frac{1}{2} \,
  e \, \frac{f^*}{\mu} \, C \, \vec{S}^{\dagger} \cdot \vec{\varepsilon}
  \ T^{\dagger}_3
\end{equation}

The vertices $\gamma N N \pi^0$, $\gamma N^* N \pi^0$ and
$\gamma \Delta N \pi^0$ of Eqs. (\ref{tkrg}), (\ref{n1440}) and (\ref{delta})
are diagrammatically represented with the same Feynman graphs as the
corresponding Kroll-Ruderman terms with charged pions in Fig. 3.

The effective Lagrangians needed for the evaluation of the Feynman diagrams
and the corresponding coupling constants are given in appendix A.
The corresponding Feynman rules are given in Appendix B of ref. \cite{yo}.

\section{Results and discussions.}
In Figs. 5 to 10 we show the total cross sections for all
the isospin channels, as well as the contribution to the total cross section
of diagrams with nucleon, $\Delta(1232)$, $N^*(1440)$, $N^*(1520)$
and $\rho(770)$ as intermediate states
(in ref. \cite{yo} we also showed differential cross sections and
invariant-mass distributions for the $\gamma p \rightarrow \pi^+ \pi^- p$
channel).
We show results up to $E_{\gamma} = 800$ $MeV$, where the new experiments at
Mainz concentrate, and we compare our results with the available experimental
data \cite{abbhhm,gialanella,carbonara,piazza,daphne}. In the errorbars of
the DAPHNE data \cite{daphne} we have plotted only the statistical errors.
The systematic errors are of the same order of magnitude.

The isospin channels with one or two charged pions in the final state are
shown in Figs. 5, 6, 8 and 9. We observe that the $\Delta(1232)$ terms
(short-dashed lines) are dominant in these isospin channels.
Essentially the $\Delta(1232)$ Kroll-Ruderman and $\Delta(1232)$
pion-pole terms (diagrams (i)
and (k) of Fig. 3) are the
more important terms.
The non resonant terms (short-dash-dotted lines) are much smaller and
they provide a small background which grows up moderately as a function of
the energy.
The $N^*(1520)$ contribution (long-dashed lines) by itself
is also small compared to the
$\Delta(1232)$ one, but it is essential to reproduce
the total cross section due to its interference with the $\Delta(1232)$
Kroll-Ruderman term
as we already remarked in ref. \cite{yo}.
This interference occurs only between the $\Delta (1232)$ Kroll-Ruderman
term and the $s$-wave part of the $N^*(1520) \Delta \pi$ contribution.
Since now part of the $N^* \rightarrow \Delta \pi$ decay is due to $d$-wave,
the interference effects are smaller than in ref. \cite{yo} where all the
$N^*(1520) \rightarrow \Delta \pi$ decay was associated to $s$-wave.
This has as a consequence a better agreement of theory with experiment
than in ref. \cite{yo}
(see Fig. 5 here in comparison with Fig. 5 of ref. \cite{yo}).
The $\rho$-terms are negligible at these energies, but they show up clearly
at higher energies (see ref. \cite{yo}).
Contributions from other terms are still smaller.

For the isospin channels with two  neutral pions in the final state
things are rather different (see Figs. 7 and 10).
In these cases a lot of terms vanish
due to the fact that the photons cannot couple to  neutral pions.
In particular, there are no Kroll-Ruderman and pion-pole terms.
Thus, the total cross section for these isospin channels is much smaller
than the cross section for the other isospin channels.
Furthermore, terms that in the other isospin channels are very small
become important in these isospin channels.

In Figs. 7 and 10 we can see that, except for the contribution from
N-intermediate states (short-dash-dotted lines), which is very small, the
other contributions are all of them relevant. Below $500$ $MeV$ we can see
that the $N^*(1440)$ (long-dash-dotted lines) dominates the  reactions.
At $500$ $MeV$ the $\Delta (1232)$ (short-dashed lines) and the $N^*(1520)$
(long-dashed lines)
start to grow up, and around $600$ $MeV$ all these diagrams
have similar strength.
At $600$ $MeV$ the $N^*(1440)$ contribution starts to fall down, and the
$N^*(1520)$ and $\Delta (1232)$  dominate the reaction.
The $N^*(1520)$ contribution peaks at $720$ $MeV$ (it is responsible of
the peak of the total cross sections) and from this energy on it falls down,
while the $\Delta (1232)$ contribution continues growing up
moderately.

It is worth noting that, in these latter cases, the $N^*(1520)$ contribution is
important by itself, and not by its interference with other terms as it
happens in the
isospin channels with one or two charged pions, and again it is essential to
reproduce the peak of the cross section around $700$ $MeV$.
To further stress the role of the $N^*(1520)$ resonance, we plot in Fig. 11
the results for the $\gamma p \rightarrow \pi^0 \pi^0 p$ cross section using
the two possible combinations of Eq. (\ref{2sol}) for the $s$- and
$d$-wave couplings.
We see that the curve $(a)$ is clearly favoured by the data, precisely the same
case which was favoured in the $\gamma p \rightarrow \pi^+ \pi^- p$ reaction.
This provides further support for our choice of amplitudes in the
$N^*(1520) \rightarrow \Delta \pi$ decay.

At this point, we should mention that in spite of the large $N^*(1440)$
decay branching ratio into $N \pi \pi$ (around 35\%), and the fact that
in the present reaction the $N^*(1440)$ is on-shell at $E_{\gamma}
\simeq 600$ $MeV$, the $N^*(1440)$ contribution to the total cross section
is very small for some of the isospin channels.
This is due to the relatively small $N^*(1440)$ coupling to photons
\cite{pdg94}, compared to the $\Delta (1232)$ or the $N^*(1520)$ resonances
(the branching ratio to $N \gamma$ is around one order of magnitude smaller
for the $N^*(1440)$ than for the $\Delta(1232)$ or the $N^*(1520)$).
However, it is important in the two-neutral pion channels in the
low energy region (below $E_{\gamma} \simeq 650 $ $MeV$) (see
long-dash-dotted line of Figs. 7 and 10).
Note again, that for $E_{\gamma} < 600$ $MeV$  the $N^*(1440)$ is the dominant
contribution, and although there are some discrepancies with de DAPHNE data
\cite{daphne} in this region, our model agrees well with the
preliminary data of the
TAPS collaboration \cite{taps} in the same region.

For the total cross sections, in Figs. 5 to 10 (solid lines)
we can see a remarkable
agreement with the experiment in some isospin channels, and some
discrepancies  in others.

For the $\gamma p \rightarrow \pi^+ \pi^- p$ channel (Fig. 5)
we reproduce quite
well the total cross section up to $E_{\gamma} = 800 MeV$ (we also
reproduce invariant mass distributions even at higher energies \cite{yo}).

For the $\gamma p \rightarrow \pi^+ \pi^0 n$ channel (Fig. 6)
we have found an
important discrepancy between our calculations and the experimental data.
It is very easy to understand qualitatively our results for the
$\gamma p \rightarrow \pi^+ \pi^0 n$ cross section from isospin coefficients.
As we have already said, the photoproduction of one or two charged pions is
dominated by the $\Delta(1232)$ Kroll-Ruderman and $\Delta(1232)$
pion-pole terms.
In the $\gamma p \rightarrow \pi^+ \pi^- p$ channel, these terms with a
$\Delta^{++}$ in the intermediate state have a factor $1$ in the amplitude
from isospin ($T^{\lambda}$), while in the $\gamma p \rightarrow \pi^+ \pi^0 p$
case with a $\Delta^0$ in the intermediate state, this factor is $\sqrt{2}/3$.
Then we have that the cross section for the $\gamma p \rightarrow \pi^+
\pi^0 n$ channel is around $4.5$ times smaller than the
$\gamma p \rightarrow \pi^+ \pi^- p$ one
(this relation is not exact because there are other terms with different
isospin coefficients that also contribute).
As we can see comparing
Figs. 5 and 6 this relation is approximately satisfied.

In order to understand these discrepancies, we have tried to look for
additional contributions for this channel.
For instance, we have calculated the contribution to the total cross
section of new diagrams with $\rho$-intermediate state decaying into $\pi \pi$
where the photon is coupled to the $N N \rho$ vertex in one case and to
the intermediate $\rho$ meson line in another case.
Note that these additional diagrams only contribute for the $\gamma p
\rightarrow \pi^+ \pi^0 n$ and $\gamma n \rightarrow \pi^- \pi^0 p$
channels, because for the others channels, the intermediate $\rho$-meson is
neutral, and does not couple to photons. However, we have
found that these contributions are very small,
and they do not modify our results.

On the other hand, we have also investigated the contribution to the total
cross section of the off-shell part of the $\Delta$ interactions with
photons and pions (see ref. \cite{off}). For this
purpose, we have evaluated diagram $(i)$ of Fig. 3, including the off-shell
part of the $\Delta$-interaction.
Then, from the off-shell parameters of the $\Delta$ interaction, we get an
additional term which is very similar to the one obtained from
diagram $(i)$ of Fig. 3 without the off-shell parameters, but it has a
multiplicative factor $(\sqrt{s_{\Delta}}-m_{\Delta})/2 m_{\Delta}$
($\sqrt{s_{\Delta}}$ is the $\Delta$ invariant mass, and $m_{\Delta}$ is the
$\Delta$ mass). This factor is zero for a $\Delta$ on-shell, and in our
range of energies, it changes between $-0.05$ and $+0.07$, and it strongly
peaks around zero (see Fig. 6 of ref. \cite{yo}).
This is so, because the pion of the $\gamma \Delta N \pi$ vertex in Fig. 3 (i)
picks up the necessary momentum to make the $\Delta$ closest to on-shell.
When we perform the actual calculations the modifications from the
consideration of these off-shell corrections are of the order of 1\%
for a broad range of the off-shell $Z$ parameter
(we have checked it for several values of $Z$ between $-1/2$ and $1/2$).
Then, we conclude that the off-shell contribution of the $\Delta(1232)$
resonance to the total cross
section is very small, and we can neglect it.

For the $\gamma p \rightarrow \pi^0 \pi^0 p$ channel (Fig. 7)
we can see a good
agreement with the experimental data above $650$ $MeV$, but our results are
below the experimental  ones for $E_{\gamma}<650 \, MeV$.
We should also mention that there are already preliminary results for this
channel from the TAPS collaboration \cite{taps}, which for $E_{\gamma}
< 600$ $MeV$ are below the DAPHNE data, and our model agrees very well
with these preliminary results in this energy region. Nevertheless, we should
wait for their final results.

For the $\gamma n \rightarrow \pi^+ \pi^- n$
we can see in Fig. 8 that we approximately
reproduce the experimental data below $650$
$MeV$, but above this energy our model overestimates the experimental data.

For the $\gamma n \rightarrow \pi^- \pi^0 p$ isospin channel (Fig. 9),
the theory is a bit below the experimental results \cite{carbonara,piazza},
but the trend of the energy dependence is well reproduced.

However, in connection with the data of refs. \cite{carbonara,piazza}
for the $\gamma n \rightarrow \pi^+ \pi^- n$ and
$\gamma n \rightarrow \pi^- \pi^0 p$ reaction, we should point out that
there is some disagreement between the results of the same
experiments for the $\gamma p \rightarrow \pi^+
\pi^- p$ channel and the experimental results
of ref. \cite{abbhhm,daphne} (see Fig. 5).
One should then be cautious and wait for new
measurements in these channels before extracting any conclusion from the
present discrepancies of our model with these data.

In Fig. 10 we show the results of our model for the
$\gamma n \rightarrow \pi^0 \pi^0 n$ channel
for which there are not yet any experimental data.

Due to the interest raised on the two pion photoproduction at threshold
\cite{taps,benmerrouche,meisner,meis,ben}, we  have also calculated the
total cross section for the $\gamma p \rightarrow \pi^+ \pi^- p$ and $\gamma
p \rightarrow \pi^0 \pi^0 p$ channels at energies near threshold.
In Figs. 12 and 13 we show our results for the total cross section for these
channels, compared to the results of ref. \cite{benmerrouche,meisner}
(in ref. \cite{benmerrouche} only the amplitude at threshold is given,
not the total cross
section. Then, we have taken that amplitude as a constant, and we have
integrated it over the three body phase space).

In Fig. 12 we see the results for the $\gamma p \rightarrow \pi^+ \pi^- p$
channel. There appear to be some discrepancies between the results of refs.
\cite{benmerrouche} and \cite{meisner} which are mostly due to the fact that
we use the constant amplitude at threshold of ref. \cite{benmerrouche}.
Indeed, if the same is done for the work of ref. \cite{meisner} the results
are very similar (see Fig. 9 of ref. \cite{meisner}). This also means that
energy dependent terms are important even very close to threshold.

Our results are above the results of refs. \cite{benmerrouche,meisner}.
We should note that in addition to
the Chiral terms which contribute at threshold, and which we have in terms of
our effective Lagrangians, there is one important contribution at threshold
which is missed in the two other approaches.
This corresponds to the diagram of Fig 3. (q) and the crossed one which is
not drown there but is actually calculated in this channel.
They correspond to $N^*(1440)$ excitation with decay into $\pi \pi$ in
$s$-wave and $N$. These terms were also shown to be important at threshold
in the $\pi N \rightarrow \pi \pi N$ reaction \cite{pidospi}.
Its relevance in that reaction has been stressed in a recent paper
\cite{bernard} where the existence of an additional energy dependent term
of unknown strength is pointed out.

The different threshold behaviour of the approaches of refs.
 \cite{benmerrouche,meisner} and the present work, apart from some
differences in the treatment of the $\Delta$ terms, is mostly due to the fact
that we include the $N^*(1440)$ resonance.
In addition there is a small contribution from the $N^*(1520)$ a few $MeV$
above threshold,
and there are also other terms which vanish at threshold but contribute
close to it.

We should also mention that in
 the region of interest to us,
$450$ $MeV \, < \, E_{\gamma} \, < \, 800$ $MeV$,
the cross section is dominated by resonance terms
which are not provided by Chiral Perturbation Theory.

In Fig. 13 we show the threshold results of the $\gamma p \rightarrow \pi^0
\pi^0 p$ channel. We observe large discrepancies between the results of
refs. \cite{benmerrouche} and \cite{meisner} which are commented in refs.
\cite{meis,ben}.
Among other differences, the results of ref. \cite{meisner} include loop
corrections which are relatively important in this channel close to
threshold \cite{meisner}.
Once again our results, which also miss the loop corrections, are
bigger than in the two approaches of refs. \cite{benmerrouche,meisner} for
the same reasons discussed above. Our results at threshold also include the
contribution of the crossed term of the $N^*(1440)$ diagram of Fig. 3(q),
which is negligible at higher energies.

Furthermore, we should also stress that, like in the $\gamma p \rightarrow
\pi^+ \pi^- p$ channel, the terms which vanish at threshold dominate the
reaction at large energies.

One limitation of our model is that we neglect unitarity. Unitarity for three
particles in the final state is a nontrivial problem \cite{amado}, and is
not the purpose of this paper to deal with it.
However, for the dominant channel ($\gamma N \rightarrow \pi \Delta
[\pi N]$), due to the fact that the $\Delta$-width is implemented in the
$\Delta$-propagators, our model approximately satisfies unitarity.
In order to estimate the errors that unitarity could introduce, we follow the
procedure of Olsson \cite{olsson} to unitarize this channel by multiplying
the $\Delta$-terms by a phase and requesting that the resulting amplitude,
after adding the background, has the phase of the $\pi N$ amplitude. The angle
$\phi$ needed for this phase, $e^{i \phi}$, is of the order of $10^0$.
We have checked, for the $\gamma p \rightarrow \pi^+ \pi^- p$
channel, that implementing this phase in the amplitude changes the results
at the level of 3\% at energies below $E_{\gamma}=800$ $MeV$.
Even increasing the angle $\phi$ to
$20^0$ the changes are of the same order of magnitude. This result is due to
the fact that, except for the already commented interference between the
$N^*(1520)$ and the $\Delta$ Kroll-Ruderman term, the interference between
the different terms of the amplitude is small.

This is only a rough approach to such a difficult problem, however it gives
hints that the unitary corrections might be small in that energy range.
This conclusion is also supported by another approximate scheme to unitarize
the Born amplitudes used in \cite{luke}. An absorptive correction
factor is used there which does not modify the cross section below
$E_{\gamma}= 800$ $MeV$, although it produces a substantial reduction at
much higher energies.
Nevertheless a rigourous treatment of this problem would be welcome.

\section{Conclusions.}
We have constructed a model for the $\gamma N \rightarrow \pi \pi N$
reaction extending the model of ref. \cite{yo}
for the $\gamma p \rightarrow \pi^+ \pi^- p$ reaction to the rest of
the isospin channels, including some improvements in the $\Delta(1232)$,
$N^*(1440)$ and $N^*(1520)$ couplings.
Our model reproduces quite well the experimental results of
refs. \cite{abbhhm,daphne,taps} below $E_{\gamma}= \ 800 \ MeV$ for the
$\gamma p \rightarrow \pi^+ \pi^- p$ and $\gamma p \rightarrow \pi^0 \pi^0 p$
isospin channels, but we have found some important discrepancies for the
$\gamma p \rightarrow \pi^+ \pi^0 n$ channel.
For the isospin channels on the neutron we have found also some discrepancies
with the old data of refs. \cite{carbonara,piazza}, but as we have already
mentioned, these experiments should be improved.

As we already observed in ref. \cite{yo} the reaction is dominated by the
$\Delta(1232)$ Kroll-Ruderman and $\Delta(1232)$ pion-pole terms,
but we get an appreciable contribution
from other terms. In particular we have shown the crucial importance of the
$N^*(1520)$ terms.
In the isospin channels with one or two charged pions in the final state, the
$N^*(1520)$ contribution is small compared with the $\Delta (1232)$ terms,
but it shows up clearly in the total cross section due to its interference
with the $\Delta(1232)$ Kroll-Ruderman term.
For the cases with two neutral pions in the final state the dominant terms
in the other isospin channels vanish, and the contribution of the $N^*(1520)$
terms is bigger than the contribution of the $\Delta (1232)$.
At energies below $E_{\gamma} = 600$ $MeV$ the $N^*(1440)$ plays an
important role in the two neutral pion channels.

One of the interesting findings was the possibility to extract information
on the coupling constants for $N^*(1520)$ decay into $\Delta \pi$.
The interference of the $N^*(1520)$ excitation term, followed by $\Delta \pi$
decay, with the dominant $\Delta(1232)$ Kroll-Ruderman term in the
$\gamma p \rightarrow \pi^+ \pi^- p$ channel allowed us to determine the
sign of the couplings for the $s$-wave and $d$-wave in the
$N^*(1520) \rightarrow \Delta \pi$ decay,
when the absolute values given by the branching ratios, and the constraint
$A_s/A_d > 0$ obtained form the $\pi N \rightarrow \pi \pi N $ reaction
were used.
We also observed that the same choice of amplitudes gave rise to a fair
agreement with the $\gamma p \rightarrow \pi^0 \pi^0 p$
channel while the other possible
choice gave rise to a cross section in sheer disagreement with the data.

While in the $\gamma n$ channels there are reasons to expect that the data
will change when new experiments are done,
the discrepancy found with the recent data \cite{daphne} of the
$\gamma p \rightarrow \pi^+ \pi^0 n$ channels is puzzling.
For the moment we cannot envisage a solution to this problem.
Alternative experiments would be in any case most welcome.

\vspace{2cm}

Acknowledgements:  Discussions with M.J. Vicente-Vacas
are warmly acknowledged.
We would also like to thank N. d'Hose, G. Audit, G. Tamas and H. Str\"oher for
multiple discussion around their preliminary data.
This work has been partially supported by \-CICYT\-
contract number AEN 93-1205.
One of us J.A. G\'{o}mez Tejedor wishes to acknowledge financial support from
the Instituci\'{o} Valenciana d'Estudis i Investigaci\'{o}.

\newpage
\appendix
\setcounter{equation}{0}
  \setcounter{section}{1}
{\centerline{\bf Appendix A}}
LAGRANGIANS.

\vskip 1.0cm
\begin{equation}                \label{lnnp}
{\cal L}_{N N \pi} =
\frac{-f}{\mu} \, \overline{\Psi} \gamma^{\mu} \gamma_5
\partial_{\mu} \vec{\phi} \cdot \vec{\tau} \, \Psi
\end{equation}
\vskip 1.4cm

\begin{equation}                \label{lnnpp}
{\cal L}_{N N \pi \pi} =-4 \pi
\frac{\lambda_1}{\mu} \, \overline{\Psi} \vec{\phi} \cdot \vec{\phi} \Psi
\end{equation}
\vskip 1.4cm

\begin{equation}                \label{ldnp}
{\cal L}_{\Delta N \pi} =
\frac{-f^*}{\mu} \Psi^{\dagger}_{\Delta} S_i (\partial_i \phi^{\lambda})
T^{\lambda} \Psi_N  \ + \ h.c.
\end{equation}
\vskip 1.4cm

\begin{equation}                \label{lddp}
{\cal L}_{\Delta \Delta \pi} =
\frac{-f_{\Delta}}{\mu} \Psi^{\dagger}_{\Delta}
S_{\Delta i} (\partial_i \phi^{\lambda}) T^{\lambda}_{\Delta}
\Psi_{\Delta} \ + \ h.c.
\end{equation}
\vskip 1.4cm

\begin{equation}                \label{lnnep}
{\cal L}_{N^* N \pi}=
\frac{-\tilde{f}}{\mu} \Psi^{\dagger}_{N^*} \sigma_i
\left( \partial_i \vec{\phi} \, \right) \cdot
\vec{\tau} \, \Psi_N \ + \ h.c.
\end{equation}
\vskip 1.4cm

\begin{equation}                \label{lnnepp}
{\cal L}_{N^* N \pi \pi}=
- C \overline{\Psi}_{N^*} \vec{\phi} \cdot \vec{\phi} \, \Psi_N \ + \ h.c.
\end{equation}
\vskip 1.4cm

\begin{equation}                \label{ldnep}
{\cal L}_{N^* \Delta \pi} = \frac{-g_{\Delta N^* \pi}}{\mu}
\Psi^{\dagger}_{\Delta} S_i (\partial_i \phi^{\lambda})
T^{\lambda} \Psi_{N^*}  \ + \ h.c.
\end{equation}
\vskip 1.4cm

\begin{equation}                \label{ldn'p}
{\cal L}_{N'^* \Delta \pi} =
i  \overline{\Psi}_{N'^*}
\left(
   \tilde{f}_{N'^* \Delta \pi} \: - \:
   \frac{\tilde{g}_{N'^* \Delta \pi}}{{\mu}^2}
   S_i^{\dagger} \partial_i \, S_j \partial_j
\right)
\phi^{\lambda}
T^{\lambda} {\Psi}_{\Delta}
 \: + \: h.c.
\end{equation}

\begin{equation}                \label{lnnf}
{\cal L}_{NN\gamma} = - e
\overline{\Psi}_N ( \gamma^{\mu} A_{\mu} -
\frac{ \chi_{_N}}{2m} \sigma^{\mu\nu} \partial_{\nu} A_{\mu} )
\Psi_N
\end{equation}
\vskip 1.4cm

\begin{equation}                \label{lppf}
{\cal L}_{\pi\pi\gamma}=
i e (\phi_+ \partial^{\mu} \phi_- \: - \:
\phi_- \partial^{\mu} \phi_+) A_{\mu}
\end{equation}
\vskip 1.4cm

\begin{equation}                \label{lnnpf}
{\cal L}_{N N \pi \gamma} =
- i {\tt q} \frac{f}{\mu} \, \overline{\Psi} \gamma^{\mu} \gamma_5 A_{\mu}
\vec{\phi} \cdot \vec{\tau} \, \Psi
\end{equation}
\vskip 1.4cm

\begin{equation}                \label{ldnpf}
{\cal L}_{\Delta N \pi \gamma} =
- i {\tt q} \frac{f^*}{\mu} \Psi^{\dagger}_{\Delta} S_i
A_i \phi^{\lambda} T^{\lambda} \Psi_N  \ + \ h.c.
\end{equation}
\vskip 1.4cm

\begin{equation}                \label{lnnepf}
{\cal L}_{N^* N \pi \gamma}=
- i {\tt q} \frac{\tilde{f}}{\mu} {\tt q} \Psi^{\dagger}_{N^*} \sigma_i
 A_i \vec{\phi} \cdot
 \vec{\tau} \, \Psi_N \ + \ h.c.
 \end{equation}
 \vskip 1.4cm

 \begin{equation}                \label{ldnf}
 {\cal L}_{\Delta N \gamma} =
 \frac{-f_{\Delta N \gamma}}{\mu}  \Psi^{\dagger}_{\Delta}
 \epsilon_{ijk} S^{\dagger}_i \left( \partial_j A_k \right) T_3
 \Psi_N \ + \ h.c.
 \end{equation}
 \vskip 1.4cm

\begin{equation}                \label{lnenf}
{\cal L}_{N^*N\gamma} =
\frac{\tilde{f}^N_{\gamma}}{\mu} \,
\overline{\Psi}_N  \sigma^{\mu\nu} \partial_{\nu} A_{\mu}
\Psi_{N^*} \  +  \  h.c.
\end{equation}
\vskip 1.4cm

\begin{equation}                \label{ln'nf}
{\cal L}_{N'^* N \gamma} = \overline{\Psi}_N
\left\{
  {\tilde{g}}_{\gamma}\vec{S} \cdot \vec{A} \: - \: i {\tilde{g}}_{\sigma}
     \left(
        \vec{\sigma} \times \vec{S} \,
     \right) \cdot
   \vec{A}
\right\} \Psi_{N'^*} \: + \: h.c.
\end{equation}
\vskip 1.4cm

\begin{equation}                \label{lnnr}
{\cal L}_{N N \rho}=
- \overline{\Psi} \left\{ G^V_{N N \rho} \, \gamma^{\mu}
\vec{\phi}^{(\rho)}_{\mu} \ - \
\frac{G^T_{N N \rho}}{2 m} \, \sigma^{\mu \nu} \partial_{\nu}
\vec{\phi}^{(\rho)}_{\mu} \right\} \cdot
\vec{\tau} \Psi
\end{equation}
\vskip 1.4cm

\begin{equation}                \label{lrpp}
{\cal L}_{\rho \pi \pi}=
- f_{\rho} \, \vec{\phi}^{(\rho)}_{\mu} \cdot
\left( \vec{\phi} \times \partial^{\mu} \vec{\phi} \, \right)
\end{equation}
\vskip 1.4cm

\begin{equation}                \label{lndr}
{\cal L}_{\Delta N \rho}=
- \sqrt{C_{\rho}} \, \frac{f^*}{\mu} \Psi^{\dagger}_{\Delta}
\epsilon_{ijk} S_i \left( \partial_j {\phi_k^{(\rho)}}^{\lambda} \right)
T^{\lambda} \Psi \ + \ h.c.
\end{equation}
\vskip 1.4cm

\begin{equation}                \label{ln'nr}
{\cal L}_{N'^* N \rho^0} =
- {\tilde{g}}_{\rho} \overline{\Psi}_N S_i \phi^{(\rho)}_i \Psi_{N'^*}
\: + \: h.c.
\end{equation}
\vskip 1.4cm

\begin{equation}                \label{lrpf}
{\cal L}_{\rho \pi \gamma}=
\frac{g_{\rho \pi \gamma}}{\mu} \epsilon^{\alpha \beta \gamma \delta}
\, \partial_{\alpha} A_{\beta} \vec{\phi} \, \partial_{\gamma}
\vec{\phi}^{(\rho)}_{\delta}
\end{equation}
\vskip 1.4cm

\begin{equation}                \label{lrppf}
{\cal L}_{\rho^0 \pi^+ \pi^- \gamma}=
e f_{\rho} \phi^{(\rho)}_{\mu}
\left( \phi_+ A^{\mu}  \phi_- \: + \: \phi_- A^{\mu} \phi_+ \right)
\end{equation}
\vskip 1.4cm

\begin{equation}                \label{lpppf}
{\cal L}_{\pi \pi \pi \gamma}=
\frac{F^{3\pi}}{6} \epsilon^{\mu \nu \alpha \beta} \epsilon^{abc}
\, A_{\mu} \, \partial_{\nu} \phi^a \, \partial_{\alpha} \phi^b \,
\partial_{\beta} \phi^c
\end{equation}
\vskip 1.4cm

\begin{equation}                \label{lpff}
{\cal L}_{\pi \gamma \gamma}=
\frac{F^{\pi}}{4} \epsilon^{\mu \nu \alpha \beta}
\phi^0 F_{\mu \nu} F_{\alpha \beta}
\end{equation}
$$
F_{\mu \nu} = \partial_{\mu} A_{\nu} \: - \: \partial_{\nu} A_{\mu}
$$
\vskip 1.4cm

In these expressions, $\Psi$, $\vec{\phi}$, $\Psi_{\Delta}$, $\Psi_{N^*}$,
$\Psi_{N'^*}$, $\vec{\phi}^{(\rho)}_{\mu}$ and $A_{\mu}$
stand for the nucleon, pion, $\Delta (1232)$, $N^*(1440)$, $N^*(1520)$,
$\rho(770)$ and photon fields, respectively; $m$ and $\mu$ are the
nucleon and pion masses.

For the $\Delta \Delta \gamma$ coupling we write directly the vertex
contribution to the Feynman rules \cite{heller} in analogy to the
$N N \gamma$ case:
\begin{equation}                \label{hddf}
-i \delta H_{\Delta \Delta \gamma} =
i \left\{ \frac{(\vec{p}+\vec{p} \, ')}{2m_{\Delta}} e_{\Delta} +
i \, \frac{e \mu_{\Delta}}{3 m} (\vec{S}_{\Delta} \times \vec{k}\, ) \right\}
\cdot \vec{\varepsilon}
\end{equation}
where $\vec{S}_{\Delta}$ is the ordinary spin matrices for a spin
$3/2$ object, $\vec{p}$ and $\vec{p}'$ stand for the initial and final delta
momentum respectively, and $\vec{k}$ for the photon momentum;
$\mu_{\Delta}$ and $e_{\Delta}$ are the magnetic moment and charge of the
delta;
$m_{\Delta}$ is the delta mass.

The coupling constants are listed below.
\vskip 7mm

{\bf Coupling constants:}
\vskip 4mm
\begin{displaymath}
\begin{array} {lrl}
f = 1     & & f^* = 2.13              \\
 & \\
\lambda_1 = 0.0075 & &                 \\
 & \\
 f_{\Delta}=0.802      & &   f_{\Delta N \gamma}=0.116             \\
 & \\
\tilde{f}=0.477              & & C=-2.29 \mu^{-1}                \\
 & \\
g_{\Delta N^*\pi}=2.07       & & e=0.3027                        \\
 & \\
\tilde{f}_{N'^* \Delta \pi}=0.911 & & \tilde{g}_{N'^* \Delta \pi}=-0.552 \\
 & \\
 \chi_{_N} = \left\{
                 \begin{array} {ll}
                 1.79  & for \ proton \\
                 -1.91 & for \ neutron
                 \end{array} \right.

 & & \tilde{f}^N_{\gamma} = \left\{
                                \begin{array} {ll}
                                 0.0173  & for \ proton \\
                                -0.0112 & for \ neutron
                                 \end{array} \right.  \\
 & \\
 \tilde{g}_{\gamma}^N = \left\{
                 \begin{array} {ll}
                 0.108  & for \ proton \\
                 -0.129 & for \ neutron
                 \end{array} \right.

 & & \tilde{g}_{\sigma}^N = \left\{
                                \begin{array} {ll}
                                 -0.049  & for \ proton \\
                                 0.0073 & for \ neutron
                                 \end{array} \right. \\

 & \\
G^V_{NN\rho}=2.9             & & G^T_{NN\rho}=18.15                     \\
 & \\
f_{\rho}=6.14                & & C_{\rho}=2                             \\
 & \\
\tilde{g}_{\rho}=0.591       & & g_{\rho\pi\gamma}=0.0378              \\
  & \\
  F^{3\pi}=0.0259 \mu^{-3}    & & F^{\pi}=0.0035 \mu^{-1} \\
  & \\
 \frac{\mu_{\Delta}}{\mu_p} = \frac{e_{\Delta}}{e}
 & & \mu_{\Delta^{++}}=1.62 \, \mu_p \\
\end{array}
\end{displaymath}

The magnetic moment of the $\Delta (1232)$, $\mu_{\Delta}$, can be calculated
in the quark model \cite{close},
with the result $\mu_{\Delta} / \mu_p = e_{\Delta}/e$.
We shall use this result, except for the $\Delta^{++}$ where we shall
use the experimental value
$\mu_{\Delta^{++}}=( 1.62 \pm 0.18)\, \mu_p$
based on the $\pi p$ bremsstrahlung
 $(\pi^{+} p \rightarrow \pi^{+}p \gamma)$ \cite{bosshard}.

\newpage
\setcounter{equation}{0}
   \setcounter{section}{2}
\centerline{\bf Appendix B}
\vskip 0.5cm
TOTAL CROSS SECTION.

The cross section for the $\gamma N \rightarrow
 \pi \pi N$ reaction
is given by

$$
\sigma = \frac{m}{\lambda^{1/2}(s, 0, m^{2})}
\frac{1}{(2\pi)^{5}}
\int \frac{d^{3} p_{4}}{2\omega_{4}}
\int \frac{d^{3} p_{5}}{2\omega_{5}}
\int d^{3}p_{2}\frac{m}{E_{2}}
$$

\begin{equation}
\delta^{4} (k + p_{1} - p_{2} - p_{4} -p_{5})
\overline{\sum_{s_i}} \sum_{s_f} |T|^{2}
\end{equation}

$$
=\frac{m^{2}}{\lambda^{1/2}(s, 0, m^{2})}
\, \frac{1}{4(2\pi)^{4}}
\int d \omega_{5} d \omega_{4} d \cos \theta_{5} d \phi_{45}
$$

\begin{equation}
\theta (1 - cos^{2} \theta_{45}) \overline{\sum_{s_i}} \sum_{s_f} |T|^{2}
\end{equation}

Where $k = (\omega,\vec{k}\,)$, $p_{1} = (E_{1},\vec{p}_{1})$,
$p_{2} =(E_{2},\vec{p}_{2})$,
$p_{4} = (\omega_{4}, \vec{p}_{4})$, $p_{5} = (\omega_{5},\vec{p}_{5})$
are the momenta
of the photon, incident proton, outgoing proton
and the outgoing pions respectively. In (B.2) $\phi_{45}$,
$\theta_{45}$ are the azimuthal and polar angles of $\vec{p}_{4}$ with
respect to $\vec{p}_{5}$ and $\theta_{5}$ is the angle of $\vec{p_{5}}$ with
the $z$ direction defined by the incident photon momentum $\vec{k}$. T is the
invariant matrix element for the reaction.
We reproduce this formula here since in Eq. (2) of ref. \cite{yo} there was a
missprint and $d \theta_5$ should have been $d cos \theta_5$,
as in Eq. (B.2) above.

\newpage
{\bf Figure captions:}

\vskip 6mm
Fig. 1: Classification of the Feynman diagrams into one point, two point and
       three point diagrams.
       Continuous straight lines: baryons.
       Dashed lines: pions.
       Wavy lines: photons and $\rho$-mesons (marked explicitly).
\vskip 6mm

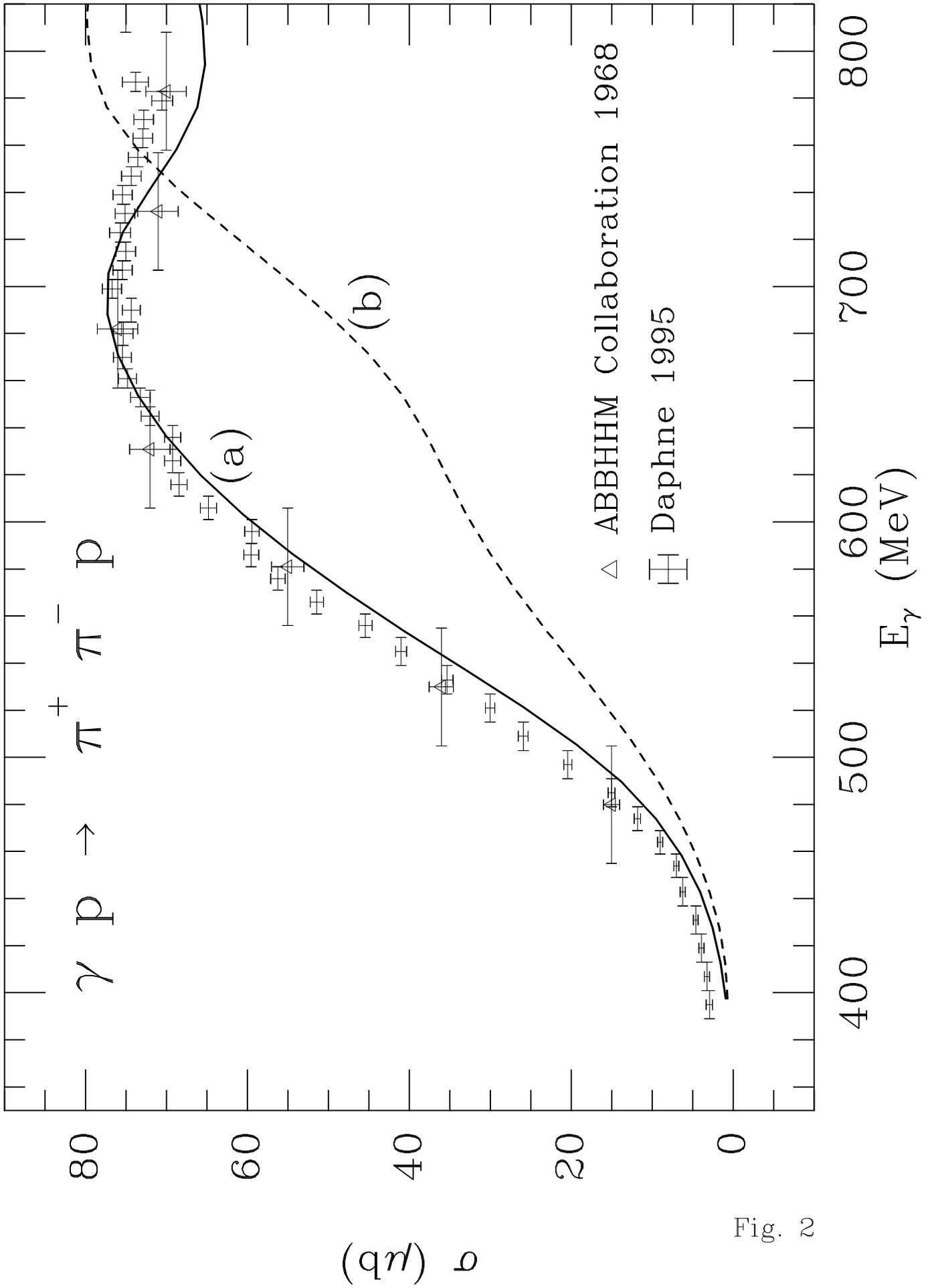
Fig. 2: Total cross section for the $\gamma p \rightarrow \pi^+ \pi^- p$
       reaction.
       { (a)} $ \tilde{f}_{N'^* \Delta \pi} = 0.911 $,
              $ \tilde{g}_{N'^* \Delta \pi} =-0.552 $;
       { (b)} $ \tilde{f}_{N'^* \Delta \pi} =-0.911 $,
              $ \tilde{g}_{N'^* \Delta \pi} = 0.552 $.
\vskip 6mm

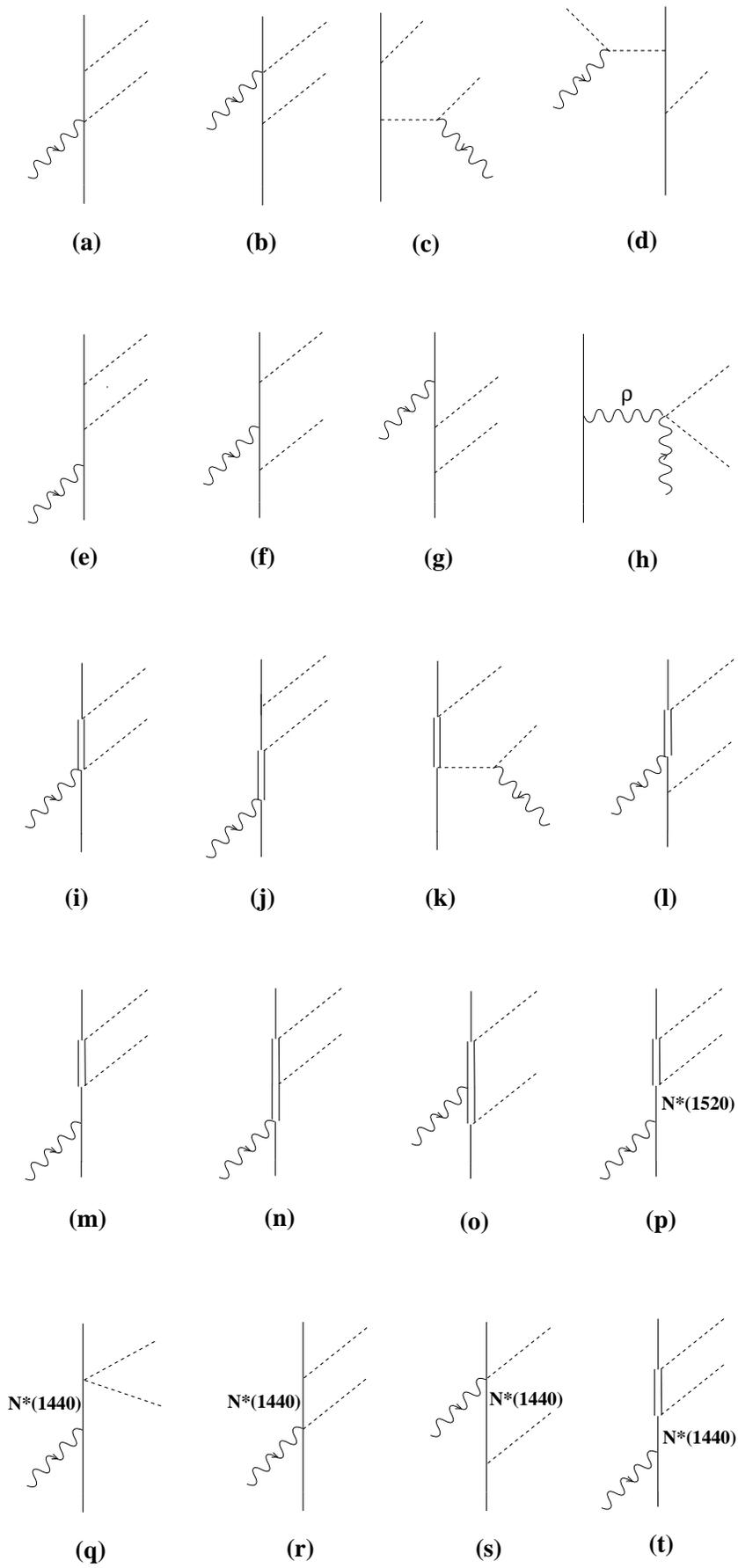
Fig. 3: Feynman diagrams for the
      $\gamma p \rightarrow \pi^+ \pi^0 n$,
      $\gamma p \rightarrow \pi^0 \pi^0 p$,
      $\gamma n \rightarrow \pi^+ \pi^- n$,
      $\gamma n \rightarrow \pi^- \pi^0 p$ and
      $\gamma n \rightarrow \pi^0 \pi^0 n$
      reactions.
\vskip 6mm

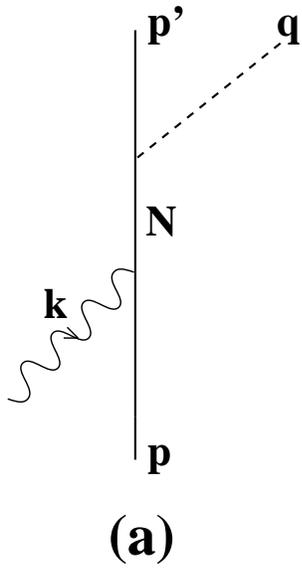
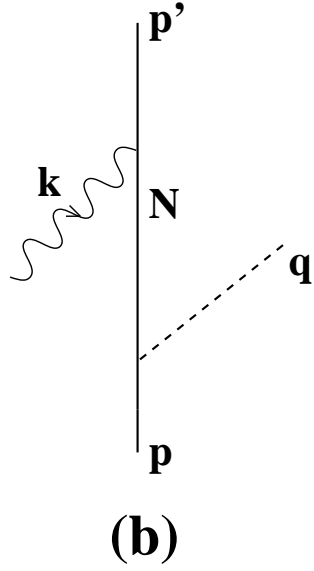
Fig. 4: Feynman diagrams for the $\gamma N \rightarrow \pi^0 N$ amplitude.
\vskip 6mm

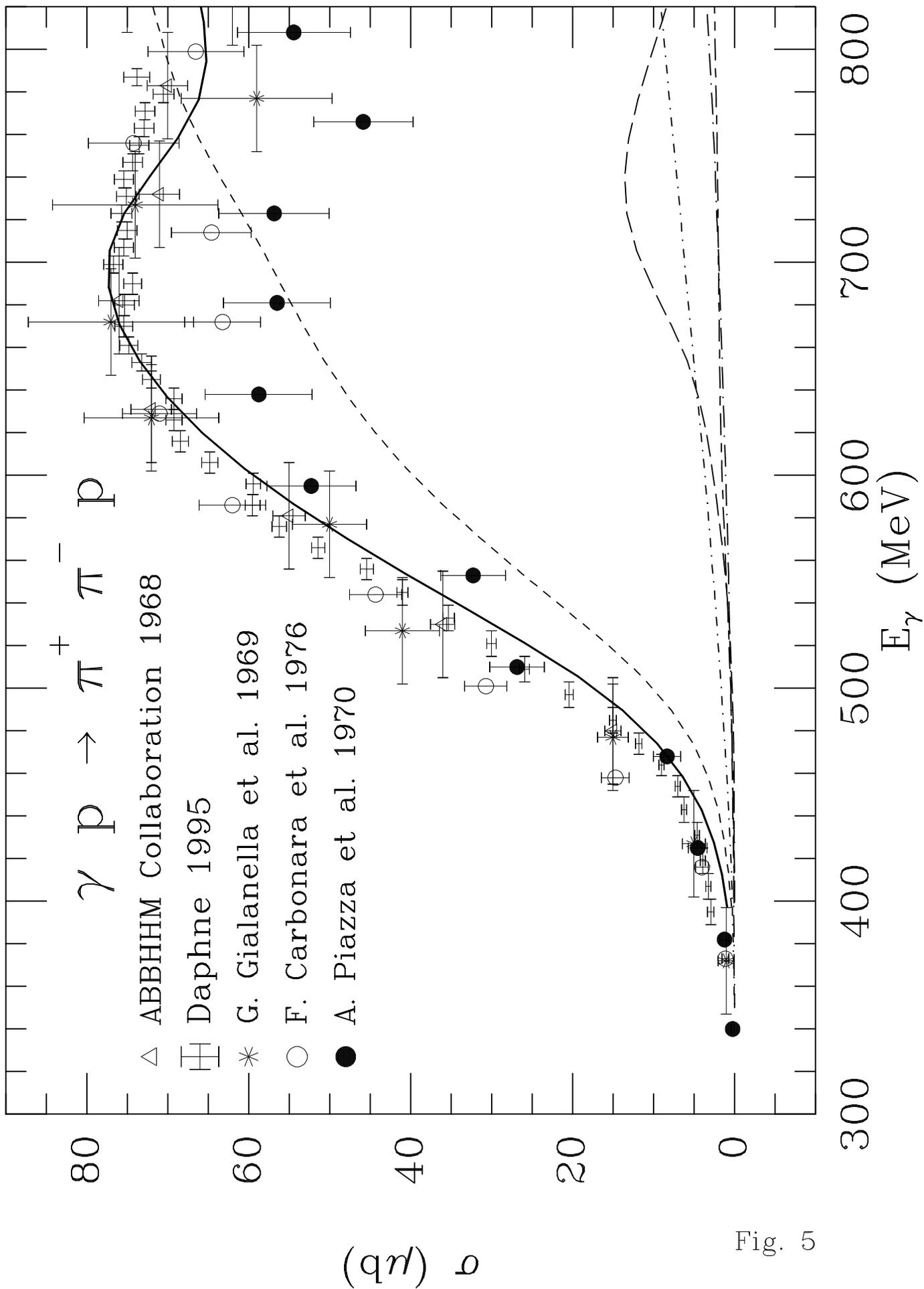
Fig. 5: Total cross section for the
      $\gamma p \rightarrow \pi^+ \pi^- p$
      reaction.
      Continuous line: total cross section.
      Short-dashed line: contribution of $\Delta (1232)$-intermediate states.
      Long-dashed line:  contribution of $N^*(1520)$-intermediate state.
      Short-dash-dotted line: contribution of $N$-intermediate states.
      Long-dash-dotted line: contribution of $\rho$-intermediate states.
      Short-dash-long-dashed line: rest of the diagrams.
      Experimental data from refs. \cite{abbhhm,gialanella,carbonara,
                                   piazza,daphne}.
\vskip 6mm

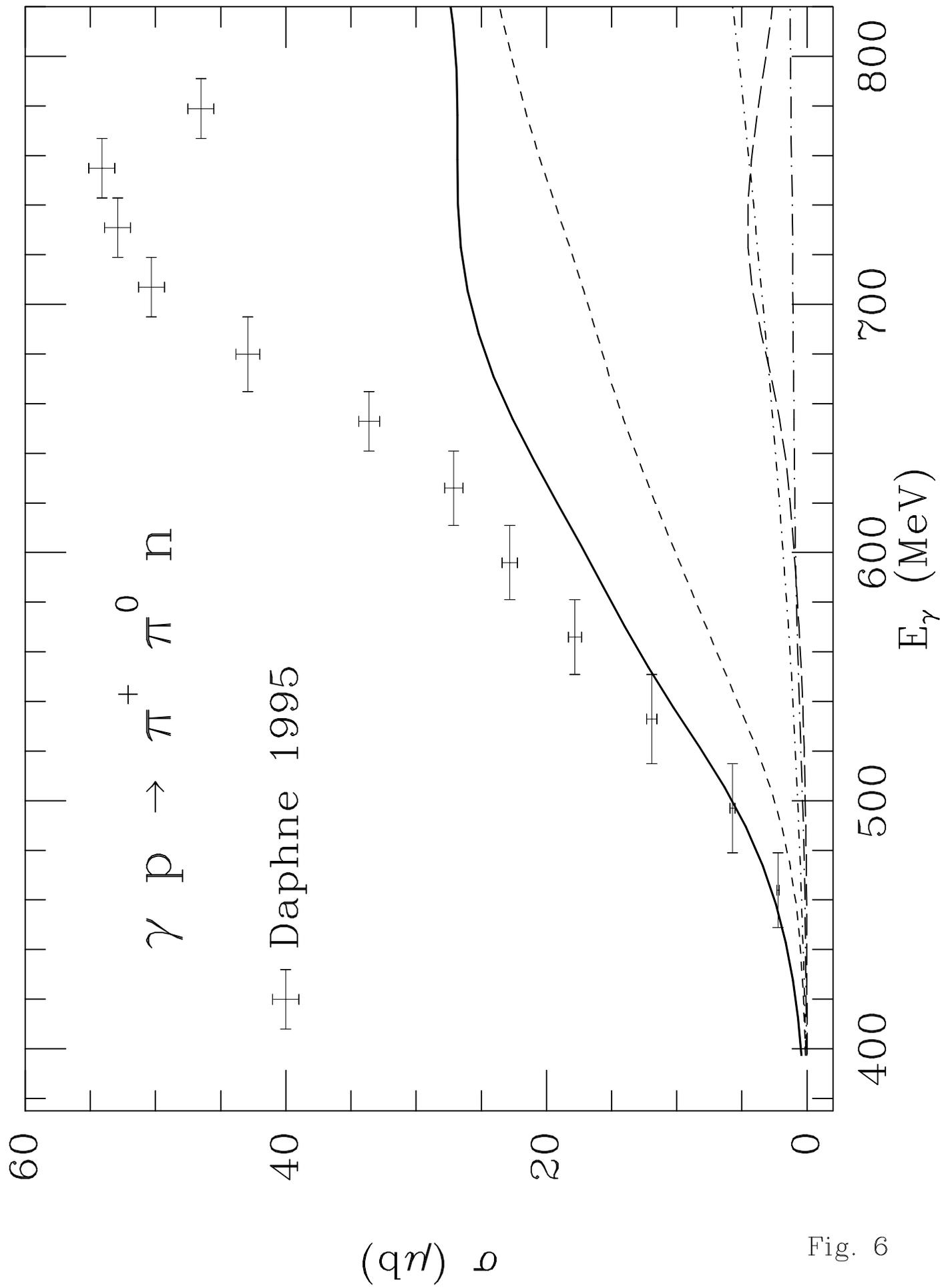
Fig. 6: Total cross section for the
      $\gamma p \rightarrow \pi^+ \pi^0 n$
      reaction.
      Continuous line: total cross section.
      Short-dashed line: contribution of $\Delta (1232)$-intermediate states
      (diagrams (i)-(o) of Fig. 3).
      Long-dashed line:  contribution of $N^*(1520)$-intermediate state
      (diagram (p) of Fig. 3).
      Short-dash-dotted line: contribution of $N$- and $\rho$-intermediate
      states (diagrams (a)-(h) of Fig. 3).
      Long-dash-dotted line: contribution of $N^*(1440)$-intermediate states
      (diagrams (q)-(t) of Fig. 3).
      Experimental data from ref. \cite{daphne}.
\vskip 6mm

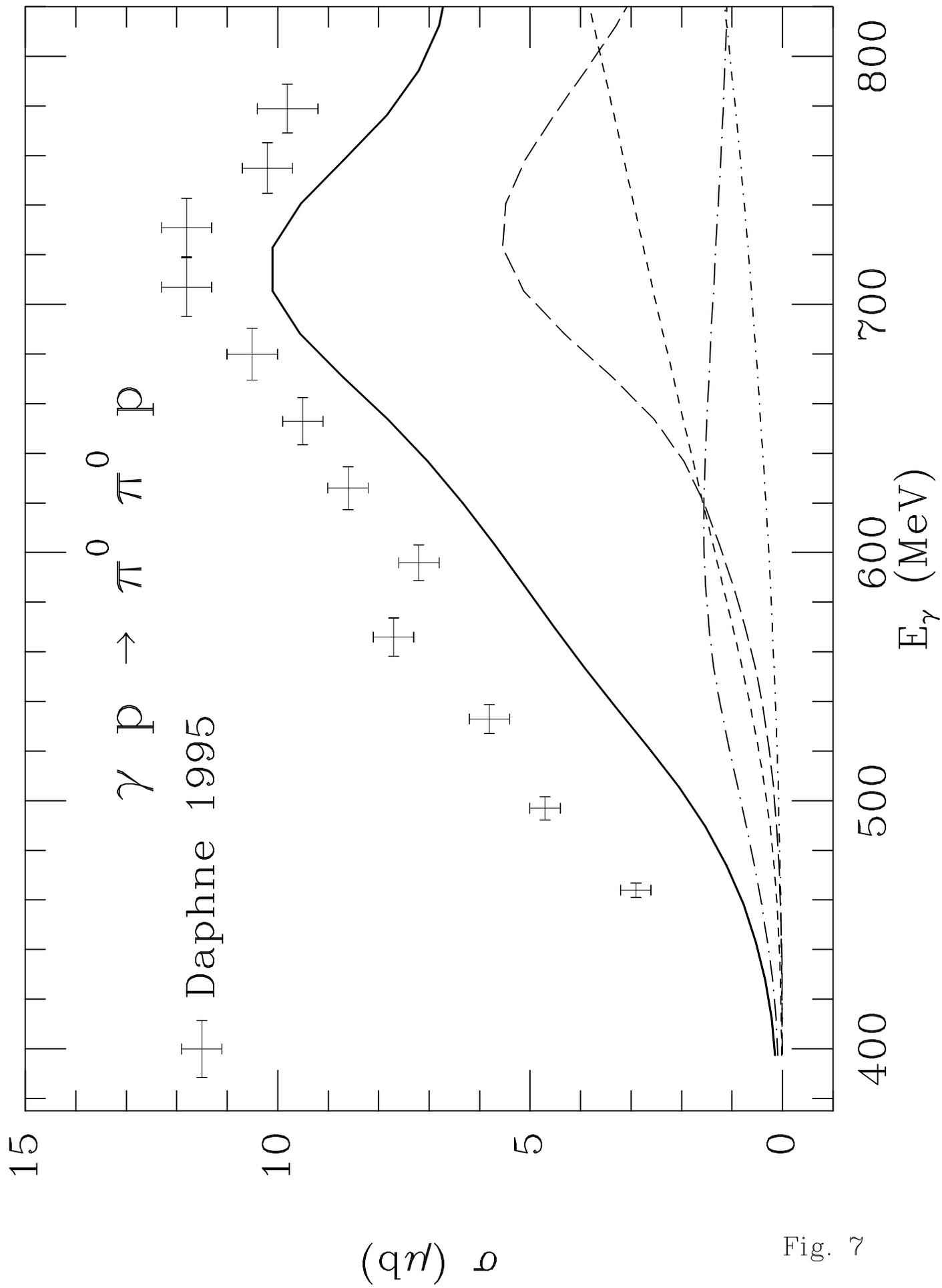
Fig. 7: Total cross section for the
      $\gamma p \rightarrow \pi^0 \pi^0 p$
      reaction.
      Continuous line: total cross section.
      Short-dashed line: contribution of $\Delta (1232)$-intermediate states
      (diagrams (i)-(o) of Fig. 3).
      Long-dashed line:  contribution of $N^*(1520)$-intermediate state
      (diagram (p) of Fig. 3).
      Short-dash-dotted line: contribution of $N$-intermediate
      states (diagrams (a)-(g) of Fig. 3).
      Long-dash-dotted line: contribution of $N^*(1440)$-intermediate states
      (diagrams (q)-(t) of Fig. 3).
      Experimental data from ref. \cite{daphne}.
\vskip 6mm

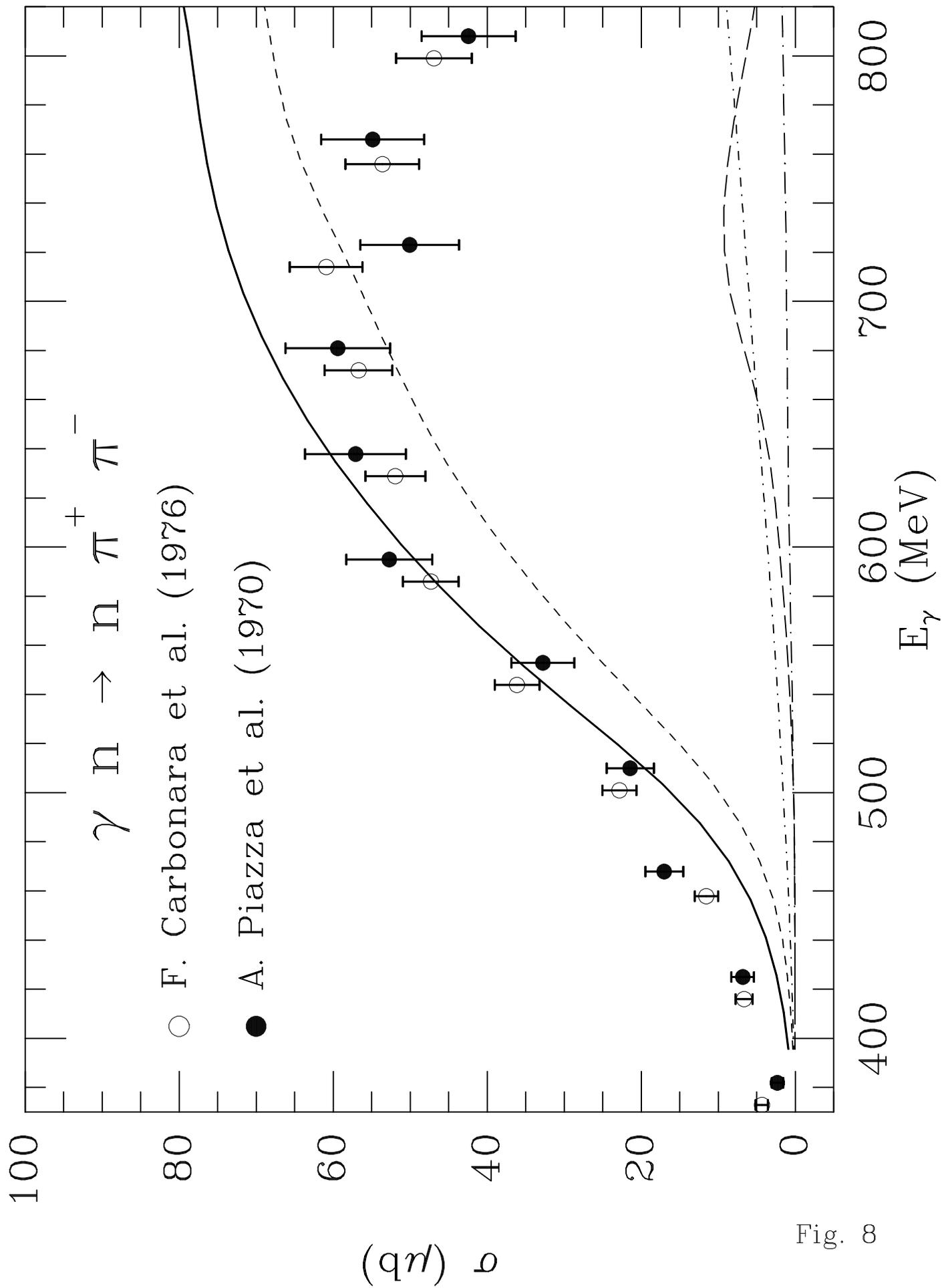
Fig. 8: Total cross section for the
      $\gamma n \rightarrow \pi^+ \pi^- n$
      reaction.
      Continuous line: total cross section.
      Short-dashed line: contribution of $\Delta (1232)$-intermediate states
      (diagrams (i)-(o) of Fig. 3).
      Long-dashed line:  contribution of $N^*(1520)$-intermediate state
      (diagram (p) of Fig. 3).
      Short-dash-dotted line: contribution of $N$- and $\rho$-intermediate
      states (diagrams (a)-(h) of Fig. 3).
      Long-dash-dotted line: contribution of $N^*(1440)$-intermediate states
      (diagrams (q)-(t) of Fig. 3).
      Experimental data from refs. \cite{carbonara,piazza}.
\vskip 6mm

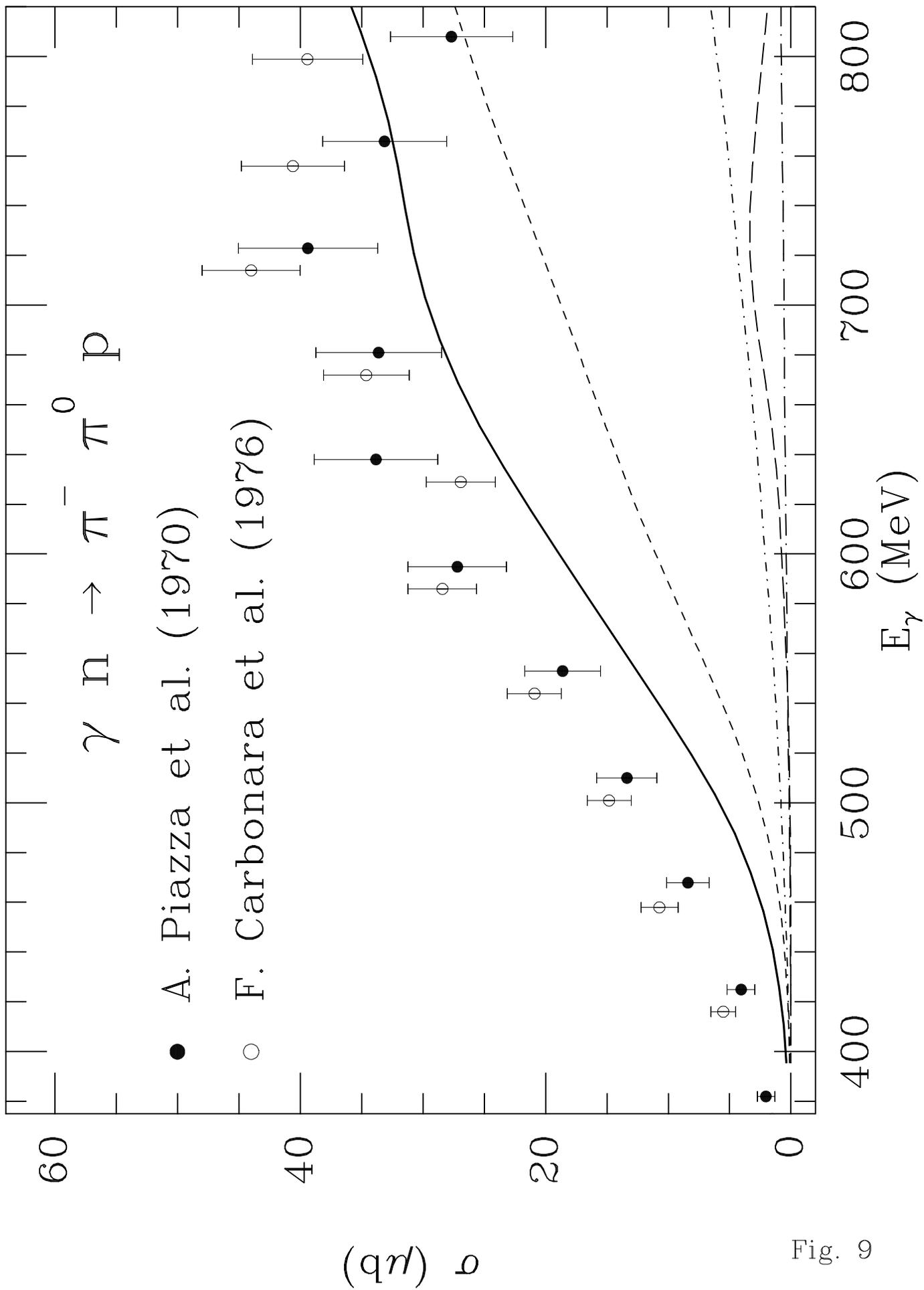
Fig. 9: Total cross section for the
      $\gamma n \rightarrow \pi^- \pi^0 p$
      reaction.
      Continuous line: total cross section.
      Short-dashed line: contribution of $\Delta (1232)$-intermediate states
      (diagrams (i)-(o) of Fig. 3).
      Long-dashed line:  contribution of $N^*(1520)$-intermediate state
      (diagram (p) of Fig. 3).
      Short-dash-dotted line: contribution of $N$- and $\rho$-intermediate
      states (diagrams (a)-(h) of Fig. 3).
      Long-dash-dotted line: contribution of $N^*(1440)$-intermediate states
      (diagrams (q)-(t) of Fig. 3).
      Experimental data from refs. \cite{carbonara,piazza}.
\vskip 6mm

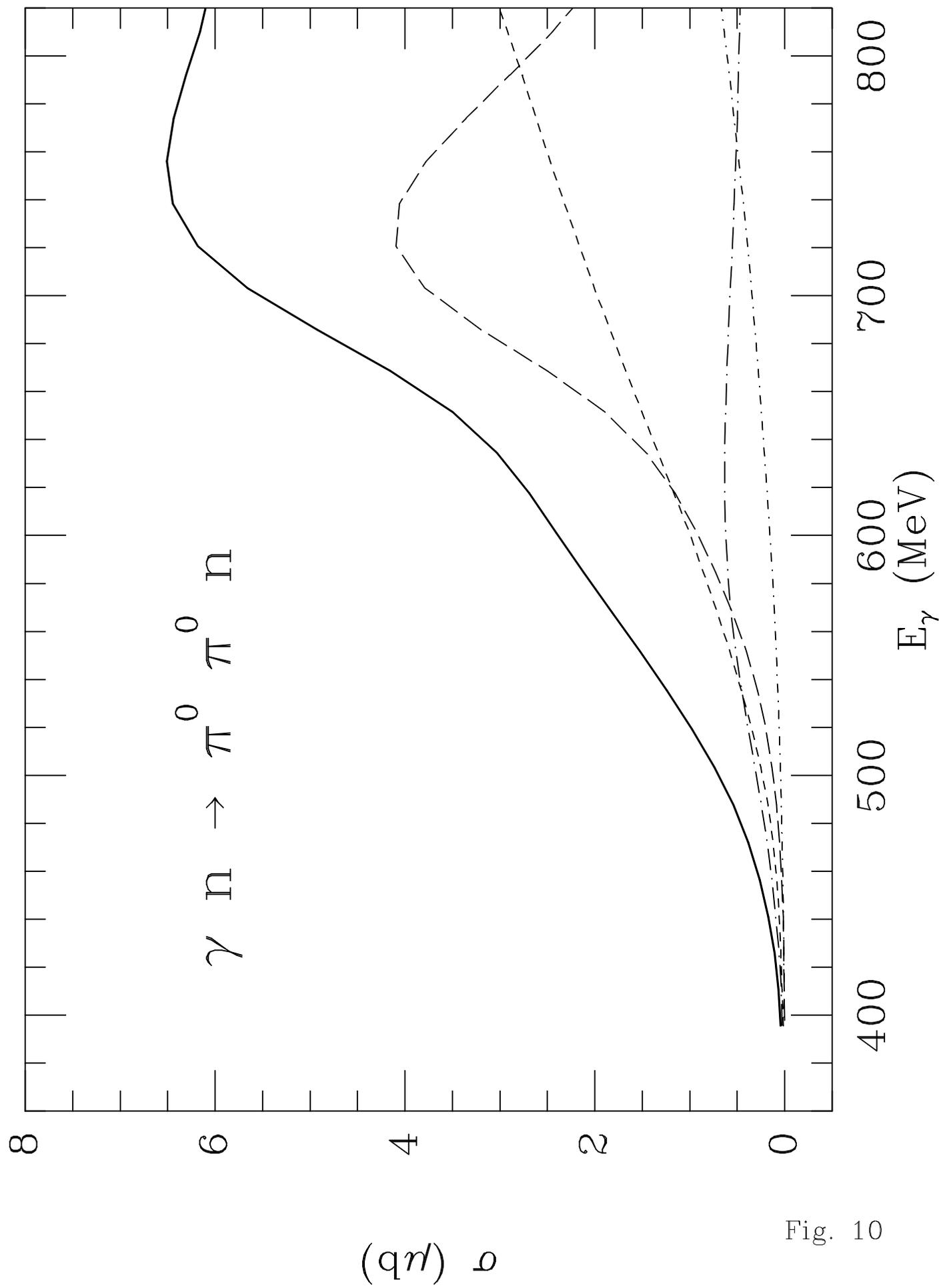
Fig. 10: Total cross section for the
      $\gamma n \rightarrow \pi^0 \pi^0 n$
      reaction.
      Continuous line: total cross section.
      Short-dashed line: contribution of $\Delta (1232)$-intermediate states
      (diagrams (i)-(o) of Fig. 3).
      Long-dashed line:  contribution of $N^*(1520)$-intermediate state
      (diagram (p) of Fig. 3).
      Short-dash-dotted line: contribution of $N$-intermediate states
      (diagrams (a)-(g) of Fig. 3).
      Long-dash-dotted line: contribution of $N^*(1440)$-intermediate states
      (diagrams (q) and (t) of Fig. 3).

\vskip 6mm
Fig. 11: Total cross section for the $\gamma p \rightarrow \pi^0 \pi^0 p$
       reaction.
       { (a)} $ \tilde{f}_{N'^* \Delta \pi} = 0.911 $,
              $ \tilde{g}_{N'^* \Delta \pi} =-0.552 $;
       { (b)} $ \tilde{f}_{N'^* \Delta \pi} =-0.911 $,
              $ \tilde{g}_{N'^* \Delta \pi} = 0.552 $.

\vskip 6mm
Fig. 12: Total cross section for the $\gamma p \rightarrow \pi^+ \pi^- p$
      reaction at energies close to threshold.
      Continuous line: results of our model.
      Long-dashed line: results of M. Benmerrouche and
      E. Tomusiak \cite{benmerrouche}.
      Short-dashed line: results of V. Bernard et al. \cite{meisner}.

\vskip 6mm
Fig. 13: Total cross section for the $\gamma p \rightarrow \pi^0 \pi^0 p$
      reaction at energies close to threshold.
      Continuous line: results of our model.
      Long-dashed line: results of M. Benmerrouche and
      E. Tomusiak \cite{benmerrouche}.
      Short-dashed line: results of V. Bernard et al. \cite{meisner}.

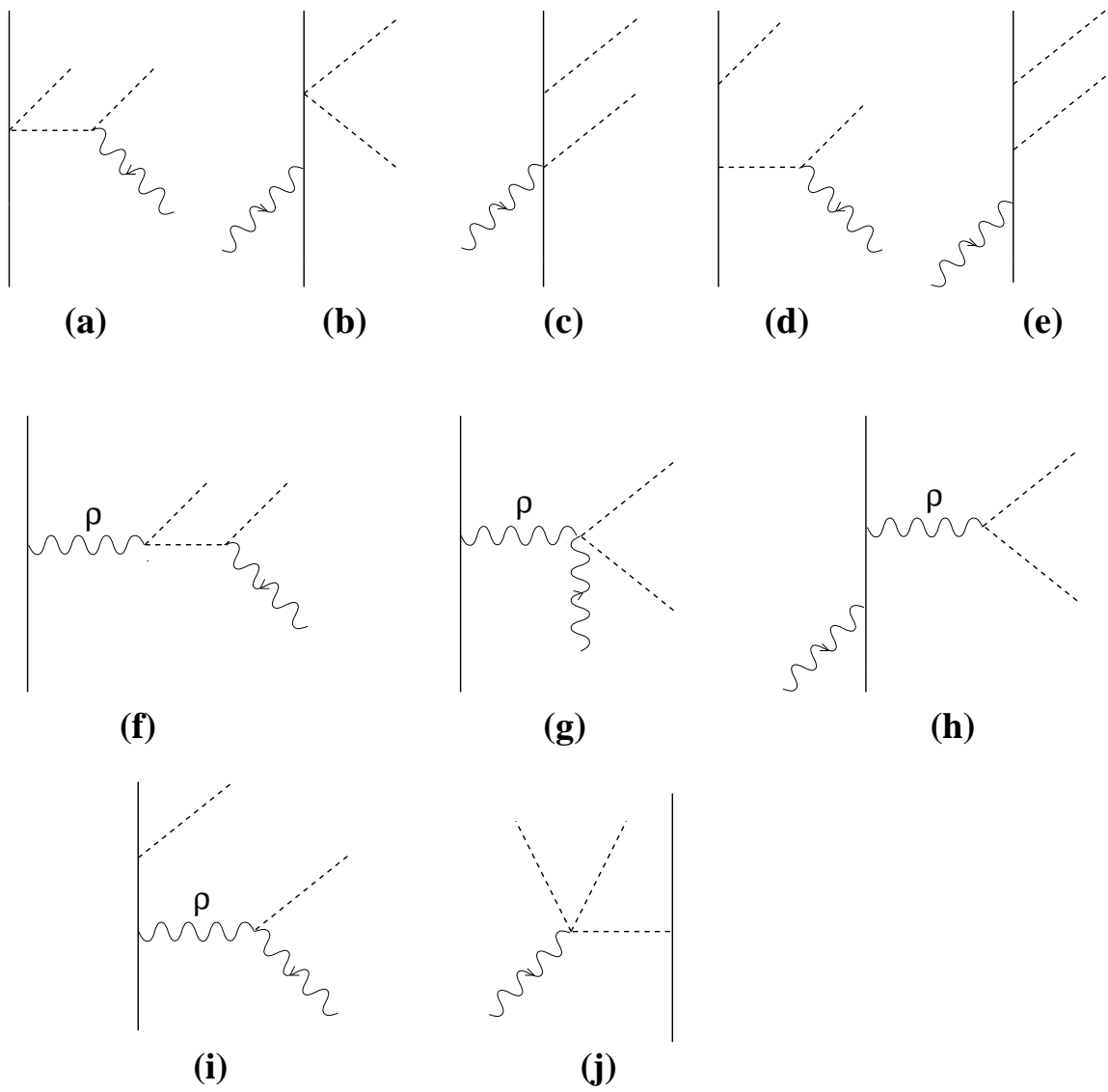
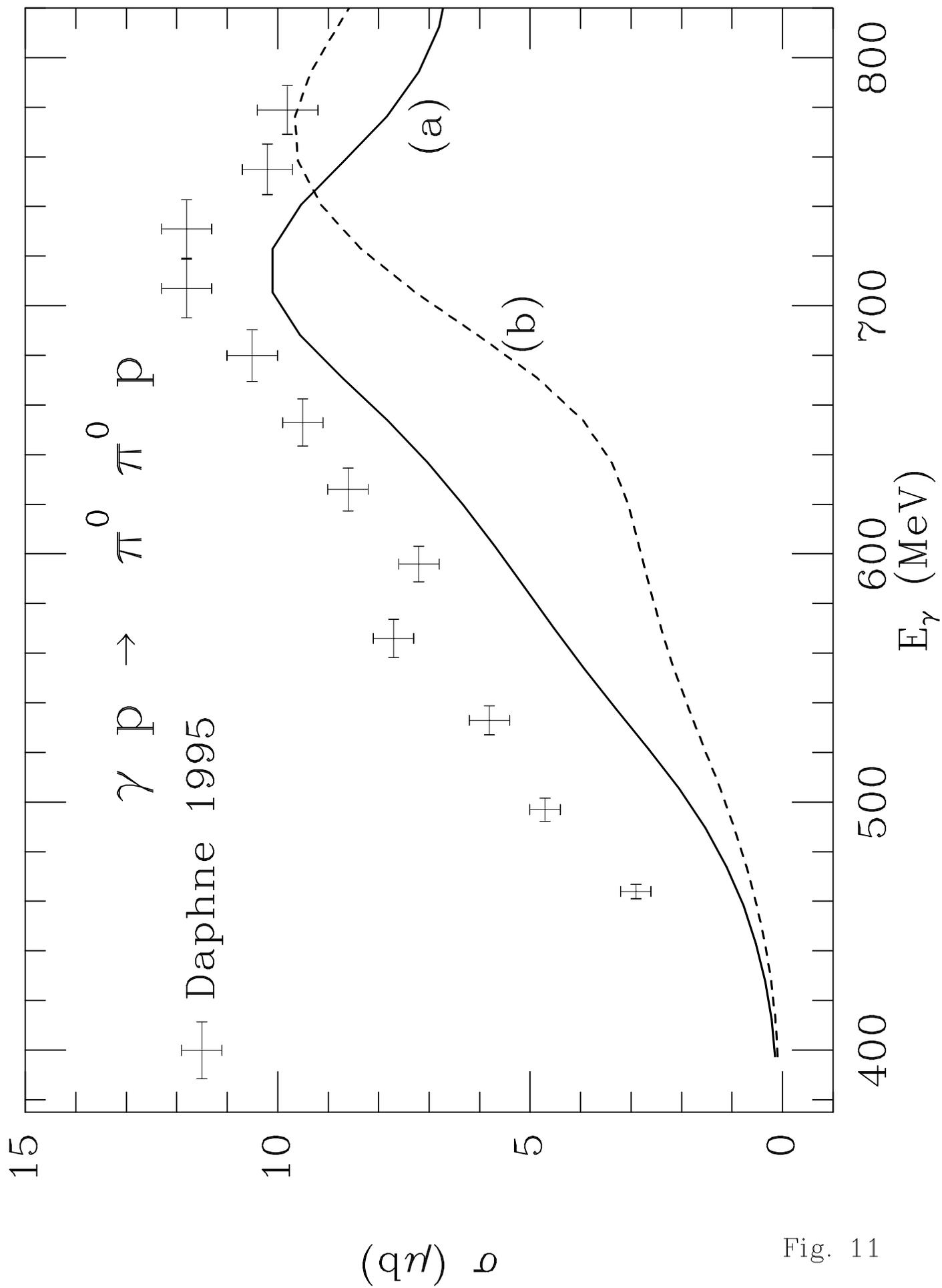
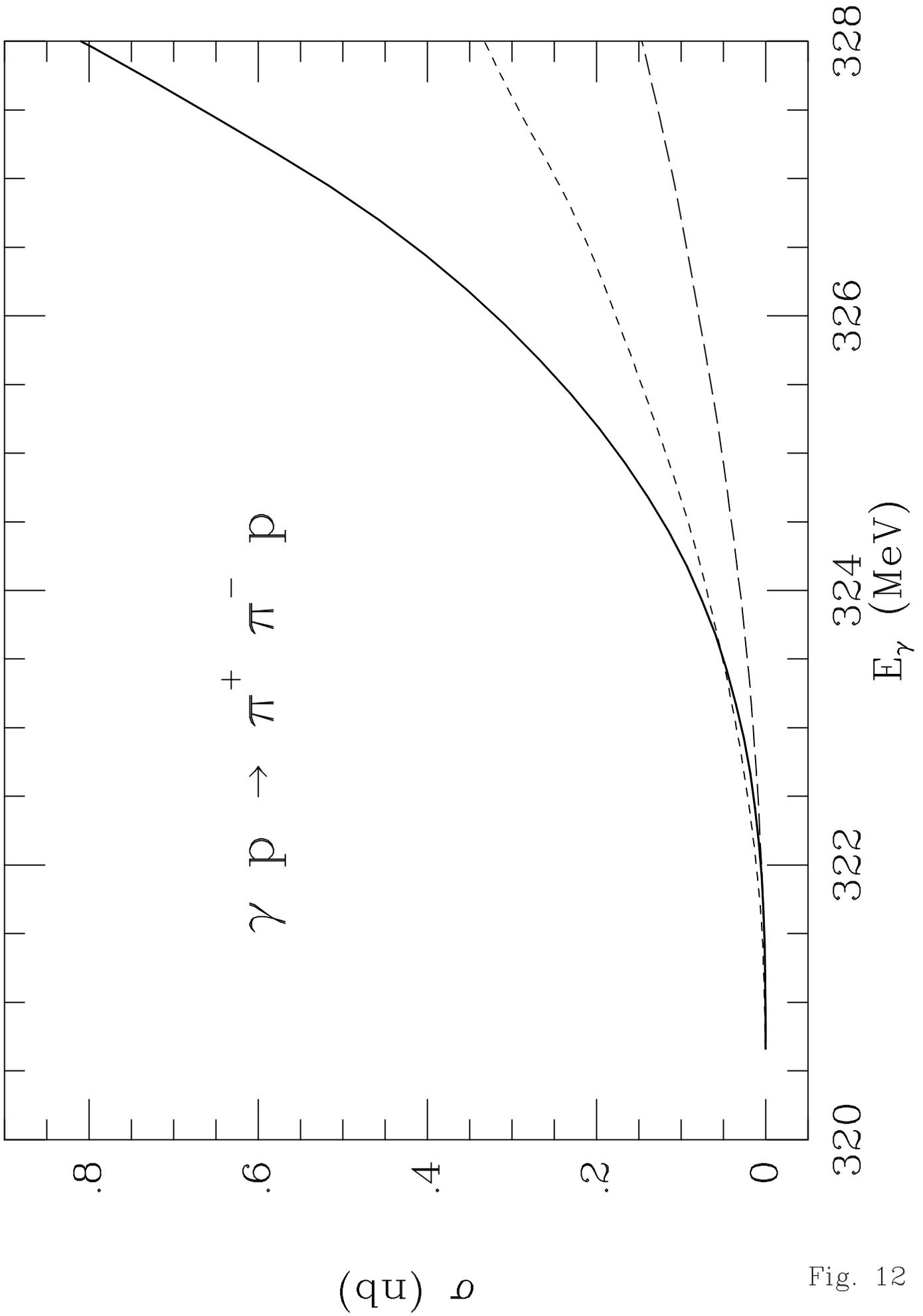
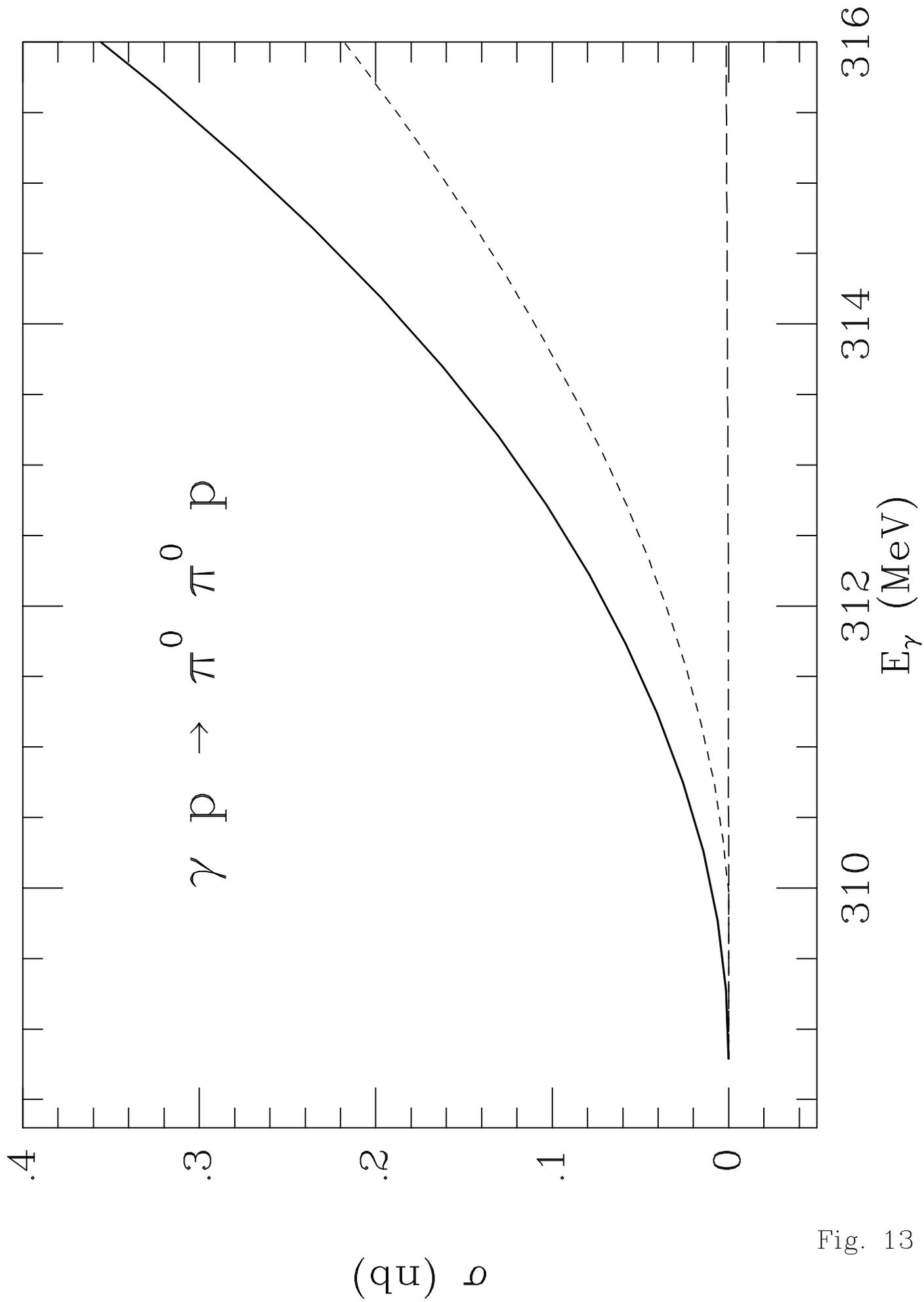
\end{document}